\documentclass[10pt]{article} 
\usepackage{epsfig}      

\newcommand{\be}{\begin{equation}}
\newcommand{\ee}{\end{equation}}
\newcommand{\ber}{\begin{eqnarray}}
\newcommand{\eer}{\end{eqnarray}}
\newcommand{\ba}{\begin{array}}
\newcommand{\ea}{\end{array}}
\newcommand{\bacc}{\begin{array}{cc}}
\newcommand{\baccc}{\begin{array}{ccc}}
\newcommand{\bacccc}{\begin{array}{cccc}}
\newcommand{\lp}{\left(}
\newcommand{\rp}{\right)}
\newcommand{\ls}{\left[}
\newcommand{\rs}{\right]}
\newcommand{\lb}{\left\{}
\newcommand{\rb}{\right\}}

\newcommand{\pivec}{\vec{\pi}}
\newcommand{\Psivec}{\vec{\Psi}}

\newcommand{\fm}{\rm \; fm}
\newcommand{\fmc}{\rm \; fm\! / \! c}
\newcommand{\mev}{\rm \; MeV}
\newcommand{\fpi}{f_\pi}
\newcommand{\mpi}{m_\pi}
\newcommand{\tauvec}{\vec{\tau}}
\newcommand{\pik}{\pi_k}

\newcommand{\tr}{{\mathrm tr}}
\newcommand{\cU}{{\cal U}}
\newcommand{\cUd}{{\cal U}^\dagger}

\newcommand{\eABMN}{\epsilon^{\alpha \beta \mu \nu}}

\newcommand{\eMNAB}{\epsilon^{\mu \nu \alpha \beta }}

\begin{document}

\title{
\bf Skyrmion-anti-Skyrmion Annihilation with $\omega$ Mesons
}

\author{M. \'A. Hal\'asz and R. D. Amado\\
{\normalsize Department of Physics and Astronomy, University of Pennsylvania}\\
{\normalsize Philadelphia, PA 19104-6396 }\\
}

\newcommand{\preprintno}{
  \normalsize UPR-902-T}

\date{\today, \preprintno}

\begin{titlepage}
\maketitle
\def\thepage{}

\begin{abstract}
We study numerically the annihilation of an $\omega$-stabilized Skyrmion and an 
anti-Skyrmion in three spatial dimensions.
To our knowledge this is a first successful simulation of Skyrmion-anti-Skyrmion
annihilation which follows through to the point where the energy is carried by
outgoing meson waves. 

We encounter instabilities similar to those encountered is earlier calculations,
but in our case these are not fatal and we are able to simulate through this 
process with a global  energy loss of less than $ 8 \%$,
and to identify robust features of the final radiation pattern.
The system passes through a singular configuration at the time of half-annihilation.
This is followed by the onset of fast oscillations which are superimposed on
the smoother process which leads to the appearence of outgoing spherical waves.

We investigate the two prominent features of this process, the 
proliferation of small, fast oscillations,
and the 
singular intermediate configuration.
We find that our equations of motion allow for a regime in which the amplitude of 
certain small perturbations increases exponentially.
This regime is similar but not identical to the situation pointed out earlier regarding
the original Skyrme model.
We argue that the singularity may be seen as the result of a pinch effect 
similar to that encountered in plasmas.
\end{abstract}

\end{titlepage}

\renewcommand{\thepage}{\arabic{page}}
\setcounter{page}{1}


\section{Introduction}

The attractive idea of representing nucleons as solitons of the effective pion
field was proposed by Skyrme long before the advent of QCD \cite{Skyrme}. 
In the context of QCD, the idea remains equally attractive, moreover 
the approach becomes exact in the large $N_c$ limit \cite{QCDskyr}. 
The Skyrme approach gives a classical, nonperturbative picture of hadrons 
in the context of low energy QCD. 
Many properties of nucleons are reobtained using a model of quantized spinning
solitons of the pion field \cite{Review}, including the static nucleon-nucleon
 and nucleon-antinucleon \cite{LPA} potential.

The main difficulty in using the Skyrme approach as a starting point for an 
effective dynamical theory of nucleons lies in the fact that all such enterprises must 
use numerical calculations to describe the dynamics of the solitons. No
 analytic solutions are known. 
One path is to study the classical dynamics of Skyrmions and use the
results to build up the quantum dynamics of nucleons based on the model of the static
nucleon as a superposition of spinning Skyrmions.
Conceptually the simplest process one might study this way is low-energy nucleon-antinucleon
annihilation since in this case there are no nucleons in the final state. 
At low energies the initial state is reasonably well described by Skyrmions,
while in the final state there are only mesons which again are 
well described by 
the effective theory \cite{Lu&RDA}.

Simulating soliton-antisoliton annihilation has proven to be a difficult
numerical problem.
Previous attempts to simulate the annihilation of a Skyrmion-anti-Skyrmion pair
\cite{Livermore,CalTech:ann,centralsc}
using the original Skyrme Lagrangian encountered numerical difficulties.
The problem has been \cite{Livermore} traced to a deviation
 from the hyperbolic nature of the
equations of motion, resulting from the contact nature of the Skyrme term.
Another possible concern is that the low-energy approximation becomes
questionable during the annihilation process when the energy tied up in the solitons 
is suddenly liberated. 
Notwithstanding, below we present a first successful attempt at such a simulation
using omega stabilization instead of the Skyrme term.  
We are able to follow the Skyrmions from incidence through annihilation 
to pion radiation. 
An interesting phenomenon involving a possible pointlike singularity 
arises in our calculation that 
might actually simplify the analysis,  however, our objective for now
is to obtain a baseline calculation using classical Skyrmion dynamics. We will
return to a detailed dicussion of the singularities in a subsequent
publication.

In this paper we present our results on Skyrmion-anti-Skyrmion annihilation using
a model \cite{SkSk,ome} which couples a $U(1)$ vector field (the $\omega$) to the
winding number of the $SU(2)$ (pion) field. This coupling replaces the Skyrme
term for stabilizing the Skyrmion. The omega stabilization is gentler than the
Skyrme term and should thus lead to less violent behavior in the simulations.
We are able to follow through the annihilation process to the point when the energy
is carried by outgoing spherical (pion and $\omega$) waves. We encountered
significant, but not fatal, numerical difficulties, 
and could ensure energy conservation to better than $8\%$ .
Our programs are set up for general initial conditions, and we will
investigate dependence on those separately. 
The results we present below refer only to the head-on annihilation process with fixed
initial velocity. To our knowledge the annihilation process 
has not been followed this far previously.

The annihilation itself, in the sense of the unwinding of the 
baryon number, takes place in a time comparable to the size of the Skyrmions. It proceeds
through an intermediary state which has a pointlike singularity, which results in the
concentration of the total energy in a very small region around the symmetry center. 
This is followed by fast oscillations which last for a time comparable to
the unwinding, and then gradually give way to outgoing
(quasi)spherical waves.

For the original Skyrme model, the Skyrme term was identified as the source of 
numerical instability \cite{Livermore}. 
Our effective Lagrangian does not have a similar local 
self-interaction term for the pion field.
However, we have encountered numerical problems similar to those expected when 
the equations are no longer purely hyperbolic, namely, the appearence of persistent
fast oscillations of small amplitude which make the simulation difficult. 
After a more careful analysis, we found that our equations of motion also allow for
non-hyperbolic solutions, (i.e., plane waves with imaginary wave number) albeit
at a higher order than the Skyrme equations. Fortunately, in our case these 
oscillations are weak enough as to not completely destroy the long-wavelength 
features. Thus our central results seem to be robust and independent of
the violent short wave length behavior. 

The most prominent feature of the annihilation process,
and probably the ultimate source of our numerical difficulties, is the fact
that close to the point of half-annihilation, when the original tips of the 
Skyrmion and the anti-Skyrmion merge, the pion field has a singular configuration.
The fields themselves are continuous, but the derivatives are singular. 
We do not yet have a complete, quantitative understanding of this phenomenon, but it is 
clear that it consists of the axial baryon current (which carries out the annihilation)
being squeezed into a very small, probably pointlike cross-section. 
This feature is localized in the $x=0$ symmetry plane (the one that separates the
Skyrmion and the anti-Skyrmion), and arises close to the moment of half-annihilation.

In the following section we define our model Lagrangian and derive the equations of motion. 
The main part of this paper is contained in Sec.3, where 
we describe our numerical results on the phenomenology of axial-symmetric annihilation and 
discuss our simulation. The first part of this section describes our
calculation, with emphasis on aspects not already discussed in \cite{SkSk}. The reader
interested only in the phenomenology of the annihilation may skip directly to Sec.3.2, 
and omit Sec.3.3 where we present numerical checks to assess the reliability of our results.
In Section 4 we investigate the stability of our equations of motion against small 
plane-wave like perturbations, in the spirit of \cite{Livermore}, and attempt to identify
signs of non-hyperbolicity in our results.
In Section 5 we discuss the 
appearence of a singularity in the pion field at the moment of half-annihilation. While 
we can understand qualitatively the mechanism that produces this phenomenon, a full understanding
will require more focused investigation.
Finally in Sec.6 we summarize our results and make a wish list of further work.

\section{The model}

Our Lagrangian consists of the nonlinear sigma model piece
\be
{\cal L}_\sigma~=~ 
\frac14 \fpi^2 \tr \lp \partial_\mu \cU \partial^\mu \cUd \rp +
\frac12 m_\pi^2 \fpi^2 \tr \lp \cU - 1 \rp~~
\ee
and the omega piece,
\be
{\cal L}_\omega~=~-\frac12 \partial_\mu \omega_\nu
( \partial^\mu \omega^\nu - \partial^\nu \omega^\mu ) + \frac12 M^2 \omega_\mu \omega^\mu
\ee
which are coupled through the baryon current,
\be
{\cal L}_{int}~=~ \frac{3 g }{2} \omega_\mu B^\mu ~~~;~~~
B^\mu~=~\frac{1}{12 \pi^2} \eMNAB \tr \lp 
\lp \cUd \partial_\nu \cU \rp
\lp \cUd \partial_\alpha \cU \rp \lp \cUd \partial_\beta \cU \rp \rp~~~.
\ee

The $SU(2)$ field $\cU$ is parameterized by the three 
real pion fields $\{\pik\}_{k=1,3} = \pivec$, or by the four 'Cartesian' components
${\bf \Psi}=\{\Psi^A\}_{A=0,3}$,
\be
\cU~=\exp \lp i \tauvec \cdot \pivec \rp~=~
\Psi_0 ~+~ i \tauvec \cdot \Psivec~=~ S^A \Psi^A ~~~~(A=0,\ldots,3)~~~.
\ee
In our previous calculation we have used the $\pi$ parameterization exclusively. 
For a detailed derivation of the equations of motion and a description of the 
corresponding numerical method we refer the reader to \cite{SkSk}. 
The more traditional method is to use the $\Psi$ parameterization. That choice
of variables has the advantage of being more transparent. Below we derive
the equations of motion in the $\Psi$ parameterization, and describe a way to employ them
in simulations without having to deal with Lagrange multipliers as dynamical
vairables. We used this approach to perform numerical calculations together
with another 
code based on the $\pi$ parameterization.
The equations of motion below are crucial for the analysis presented
in Section 4.

In the $\Psi$ parameterization the baryon current is
\be
\label{barydens}
B^\mu~=~\frac{1}{12 \pi^2} \eMNAB \epsilon^{ABCD} \Psi^A \partial_\nu \Psi^B
\partial_\alpha \Psi^C \partial_\beta \Psi^D~~,
\ee
and the nonlinear $\sigma$ piece is given by
\be
{\cal L}_\sigma~=~\frac12 f_\pi^2 \partial_\mu \Psi^A \partial^\mu \Psi^A + 
\frac12 m_\pi^2 \fpi^2 \lp \Psi^0 - 1 \rp~~.
\ee
The $\Psi^A$'s are subject to the chiral constraint which is equivalent to the
unitarity condition on the $\cal U$'s:
\be
\label{chiral}
\cU \cUd ~=~{\bf 1} ~~~\rightarrow~~~ \Psi^A \Psi^A~=~1~~~.
\ee
To get meaningful equations of motion, we must 
impose this constraint separately on the components
$\Psi^A$. 
One way is to introduce a Lagrange multiplier $\lambda$, adding a term 
$\frac{\lambda}{2} \lp  \Psi^A \Psi^A - 1 \rp$
to the Lagrangian. The physical meaning of the multiplier is similar to that of 'reaction' forces
in mechanics, which enforce constraints without performing any work. 
Obviously, the equation of motion for $\lambda$ is just the constraint equation (\ref{chiral}), 
which now has to be solved along with the other equations of motion.

Our task is to solve formally the ordinary equations of motion for the dynamical 
variables and their derivatives. These will contain $\lambda$. Then, one uses the result and
the constraint equation to solve for $\lambda$. Below, this will
 turn out to be straightforward.

To obtain the equations of motion for $\Psi^A$, start 
with the derivatives of ${\cal L}_\sigma~+~{\cal L}_{\lambda}$:
\ber
\nonumber
\frac{\delta ({\cal L}_\sigma + {\cal L}_{\lambda})}{\delta (\Psi^A)}~=~m_\pi^2  f_\pi^2 \delta^{A0} \Psi^0
~+~\lambda \Psi^A ~~;~~
\frac{\delta ({\cal L}_\sigma + {\cal L}_{\lambda})}{\delta (\partial_\mu \Psi^A)}~=~
f_\pi^2 \partial^\mu \Psi^A ~~;~~
\partial_\mu \frac{\delta ({\cal L}_\sigma + {\cal L}_{\lambda})}{\delta (\partial_\mu \Psi^A)}~=~
f_\pi^2 \partial_\mu \partial^\mu \Psi^A ~~.
\eer
The derivatives of ${\cal L}_{int}$ are:
\ber
\nonumber
\frac{\delta {\cal L}_{int}}{\delta (\Psi^A)}~&=&~\frac{g}{8 \pi^2} 
\eMNAB \omega_\mu \epsilon^{ABCD} \partial_\nu \Psi^B  \partial_\alpha \Psi^C  \partial_\beta \Psi^D
\nonumber \\
\partial_\nu \frac{\delta {\cal L}_{int}}{\delta (\partial_\nu \Psi^B)}~&=&~\frac{3 g}{8 \pi^2} \lb
\eMNAB \omega_\mu \epsilon^{ABCD} \partial_\nu \Psi^A  \partial_\alpha \Psi^C  \partial_\beta \Psi^D~+~
\eMNAB \partial_\nu \omega_\mu \epsilon^{ABCD} \Psi^A  \partial_\alpha \Psi^C  \partial_\beta \Psi^D~\rb~.
\nonumber
\eer
The $\Psi^A$ equation of motion is:
\ber
\label{psi_eom}
f_\pi^2 \partial_\mu \partial^\mu \Psi^A ~&=&~ \frac{3 g}{8 \pi^2} \eABMN \epsilon^{ABCD}
\lb  \partial_\nu \omega_\mu \Psi^B \partial_\alpha \Psi^C \partial_\beta \Psi^D ~+~
\frac43 \omega_\mu \partial_\nu \Psi^B \partial_\alpha \Psi^C \partial_\beta \Psi^D \rb ~+~
\nonumber\\&~&~~~+~
m_\pi^2 f_\pi^2 \delta^{A0} \Psi^0~+~ \lambda \Psi^A~~.
\eer

We can now proceed to eliminating $\lambda$. 
The chiral condition means that ${\bf \Psi}=\lb \Psi^A \rb$ is a four-dimensional unit vector.
This leads to constraints of its derivatives. The first derivative of $\bf{\Psi}$ with 
respect to any one of the four cordinates must be perpendicular to $\bf{\Psi}$, and so on:
\ber
\nonumber
\Psi^A \Psi^A~=~1 ~~\rightarrow~~ \partial_\mu \Psi^A \Psi^A ~=~0 ~~\rightarrow~~
\sum\limits_A \lp \partial^2_\mu \Psi^A \rp \Psi^A + \sum\limits_A \lp \partial \Psi^A \rp^2 ~=~ 0~~,
\eer
or, after summing over $\mu$:
\ber
\label{chiral2}
\partial_\mu \partial^\mu \Psi^A \Psi^A ~+~ \partial_\mu \Psi^A \partial^\mu \Psi^A ~=~ 0~~.
\eer
Geometrically this means that the component of $\partial_\mu \partial^\mu \bf{\Psi}$ parallel to 
$\bf{\Psi}$
is not a dynamical variable, but rather it is determined by the constraint. The corresponding 
part of the equation of motion (\ref{psi_eom}) therefore carries no information about $\bf{\Psi}$.
Instead, it tells us what $\lambda$ should be in order to make sure (\ref{chiral2}) and thus (\ref{chiral})
are verified.

Now we just project (\ref{psi_eom}) onto $\bf{\Psi}$ and solve for $\lambda$:
\ber
\lambda~&=&~\fpi^2 \Psi^A \partial_\mu \partial^\mu \Psi^A~-~
\frac{g}{2 \pi^2} \eMNAB \epsilon^{ABCD} \omega_\mu \Psi^A \partial_\nu \Psi^B \partial_\alpha \Psi^C 
\partial_\alpha \Psi^D~-~m_\pi^2 \fpi^2 ( \Psi^0 )^2 \nonumber \\
&=&~- ~\fpi^2 \partial_\mu \Psi^A \partial^\mu \Psi^A~-~
\frac{g}{2 \pi^2} \eMNAB \epsilon^{ABCD} \omega_\mu \Psi^A \partial_\nu \Psi^B \partial_\alpha \Psi^C 
\partial_\alpha \Psi^D~-~m_\pi^2 \fpi^2 ( \Psi^0 )^2~~.
\eer
From the second form it is clear that we have indeed solved for $\lambda$. Remember that we have
a set of second-order partial differential equations. An initial condition specifies all the fields and their first
derivatives, so the right-hand side of the second equation contains only known quatities. We can now replace this expression for $\lambda$ into the full equation of motion. 

One more observation is necessary. Since ${\bf \Psi }$ is a unit vector,
all the $\partial_\mu{\bf \Psi} $'s are perpendicular to it. Consider the quantity
\be
\eMNAB \epsilon^{ABCD} \partial_\nu \Psi^B \partial_\alpha \Psi^C \partial_\beta \Psi^D~~.
\ee
Each nonzero term in the sum over Lorentz indices is perpendicular (in isospin space) to 
three distinct vectors $\partial_\nu {\bf \Psi} ,\ldots$, all of which are perpendicular to 
${\bf \Psi }$.
(Contracting with any of them would give zero because of the $\epsilon^{ABCD}$). Furthermore,
the three vectors have to be linearly independent in order to give a nonzero contribution.
But there are four mutually perpendicular directions altogether in this space, therefore
the above quantity is necessarily parallel to $\bf{\Psi}$:
\be
\eMNAB \epsilon^{ABCD} \partial_\nu \Psi^B \partial_\alpha \Psi^C \partial_\beta \Psi^D~=
\Psi^A \eMNAB \epsilon^{EBCD} \Psi^E \partial_\nu \Psi^B \partial_\alpha \Psi^C \partial_\beta \Psi^D~.
\ee

Now we are ready to replace $\lambda$ in the equation of motion, and we obtain
\ber
f_\pi^2 \ls ( \partial_\mu \partial^\mu \Psi^A ~-~ \Psi^A (\Psi^E \partial_\mu \partial^\mu \Psi^E) \rs
~&=&~ 
\frac{3 g}{8 \pi^2} \eABMN \epsilon^{ABCD}
\partial_\nu \omega_\mu \Psi^B \partial_\alpha \Psi^C \partial_\beta \Psi^D \nonumber \\
&~& ~~~+~
m_\pi^2 f_\pi^2 \Psi^0 \lp \delta^{A0} - \Psi^A \Psi^0 \rp~~.
\eer
The left-hand side is just the piece of $\partial_\mu \partial^\mu \bf{\Psi}$ which is perpendicular
to $\bf{\Psi}$, 
$\lp\partial_\mu \partial^\mu \bf{\Psi}\rp_\perp = 
\partial_\mu \partial^\mu \bf{\Psi} - 
\bf{\Psi} \lp  \bf{\Psi} \cdot\partial_\mu \partial^\mu \bf{\Psi} \rp$. 
So, the equation of motion is written compactly:
\ber
\lp \partial_\mu \partial^\mu \Psi^A \rp_\perp
~-~ m_\pi^2 \Psi^0 \lp \delta^{A0} - \Psi^0 \Psi^A \rp~=~
\frac{3 g}{8 \fpi^2 \pi^2 } \eABMN \epsilon^{ABCD}
\partial_\nu \omega_\mu \Psi^B \partial_\alpha \Psi^C \partial_\beta \Psi^D~~.
\eer

It is easy to derive the $\omega$ equation of motion,
\ber
\label{ome_eom}
\partial^\nu \partial_\nu \omega^\mu ~&=&~
\partial^\mu \partial_\nu \omega^\nu ~-~ M^2 \omega^\mu 
~+~ \frac{g}{8 \pi^2}  \eMNAB \epsilon^{ABCD} \partial_\nu \Psi^A  \Psi^B
\partial_\alpha \Psi^C \partial_\beta \Psi^D~~.
\eer

This completes the set of equations of motion in the $\Psi$ parameterization.
They have the virtue of being more transparent than the ones based on the $\pi$ fields.

\section{Numerical results}

This section contains the central result of the paper. The reader interested only in the
physics of annihilation as we understand it may skip the technical part of the first subsection
and the third subsection in its entirety.

The purpose of the work reported here is to establish to which extent a three-dimensional
numerical simulation of Skyrmion-anti-Skyrmion annihilation can be performed in the 
$\omega$-stabilized model. We will see below that this simulation can indeed be performed
successfully. 

In the following subsection
we describe our choice of parameters for the problem. 
These choices were driven by the fact that to our knowledge this is the first 
successful calculation that follows through the annihilation of a stable three-dimensional 
soliton and its antisoliton\footnote{
A very interesting recent calculation of scattering of metastable baby Skyrmions 
\cite{Krishna} reported results on annihilation as well.
}, therefore our focus was on performing the most numerically accessible calculation 
 which has the important qualitative features of the general case.

The second subsection contains a detailed description of the phenomenology of the
central annihilation calculation. The annihilation proceeds through a 
sequence of very fast-varying 
intermediate configurations where most of the total energy is concentrated in a region
of about half the linear size of one Skyrmion. This is followed by outgoing waves, which
we followed for about $5 \fmc$. Fast oscillations of small amplitude accompany the radiation
phase.

The third subsection reports numerical checks on the reliability of our calculation. 
We compare results for the same calculation performed with different lattice spacings
and show that while there are fluctuations, the macroscopic features, such as the 
time dependence of energy flow and its angular distribution, are robust.

\subsection{Simulation}

The main challenge of this calculation lies in coping with the fast spatial
and temporal variations of the field in the annihilation process.
These require finer spatial grids and smaller timesteps than smoother processes like
soliton scattering in order to avoid increasing numerical error which ultimately can 
make a calculation meaningless. Our choice of parameters is motivated by the desire to 
reduce this problem as much as possible.

We chose the parameters close to those used in our previous work on scattering
\cite{SkSk}, but took a smaller mass for the vector field. This choice leads to softer 
dynamics, but dynamics which do not differ qualitatively from those for a physical mass.
Our choice of parameters is therefore the same as in \cite{SkSk}, with $f_\pi = 64 \mev $,
$m_\pi=139 \mev$, $g = \frac{m_\omega}{f_\pi \sqrt{2}}$, but $m_\omega=385 \mev$ .
The Skyrmions in our case are slightly smaller in spatial size 
($1.1 \fm$ versus $1.4 \fm$ in diameter), and have a mass of $650 \mev$.

We consider the special case of head-on annihilation.
The original direction of motion is along the $x$ axis. The Skyrmion and the anti-Skyrmion
are located symmetrically on opposite sides of the $x=0$ plane, with their centers
on the $x$ axis. The Skyrmion is a standard hedgehog field configuration centered at $x=-1.5 \fm$.
The anti-Skyrmion is obtained by charge conjugating 
($ \lp \Psi^0,\Psi^1,\Psi^2,\Psi^3 \rp \rightarrow \lp \Psi^0,-\Psi^1,-\Psi^2,-\Psi^3 \rp$) 
a hedgehog configuration centered at 
$x=1.5 \fm$ and then performing a grooming of $180^o$ around the $1$ or $x$ axis 
($ \lp \Psi^0,\Psi^1,\Psi^2,\Psi^3 \rp \rightarrow \lp \Psi^0,-\Psi^1,\Psi^2,\Psi^3 \rp$
altogether, the anti-Skyrmion is obtained from the Skyrmion by changing the sign of the 
$x$ or $\Psi^1$ component). This choice of grooming corresponds to the most attractive 
interaction between Skyrmion and anti-Skyrmion.

The central annihilation problem has an additional axial symmetry compared to the general case. 
However, our three-dimensional codes take only partial advantage of the axial symmetry.
We have  full three-dimensional programs which we plan to use to perform a 
sweep of many initial conditions. 
We expect that most of the features of off-center annihilation (not head on) are 
encountered in the present 
setup, since the relative position and orientation of the solitons in the general case 
is very similar to that in central annihilation\footnote{We in fact did perform a few off-center
runs.}.

We performed simulations of Skyrmion-anti-Skyrmion annihilation using both the algorithm
presented in \cite{SkSk} and a similar calculation, based on the equations of motion
from the previous section. Below we describe the latter in more detail. The only major
modification of the first calculation compared to \cite{SkSk} is the treatment of the
$\omega$ field which is similar to that described below. The two calculations give virtually 
identical results in the smooth regime, with small differences in the violent regime.
In the next subsection we present the results of a calculation using the $\Psi$ scheme.
The scaling analysis runs in subsection 3.2 use the $\pi$ scheme. 
Since the $\Psi$ scheme is more transparent, that is used for the study
of the small oscillations and of the singularity in the respective sections.

One important modification compared to \cite{SkSk}, which also affects the $\pi$ scheme, 
refers to the implementation of the
gauge fields. Taking the four-divergence of the $\omega$ equation of motion, 
together with baryon current conservation leads to 
\ber
M^2 \partial_\mu \omega^\mu ~=~ 0~~.
\eer
In other words, the nonzero mass breaks gauge symmetry enforcing the Lorentz gauge. 
We can drop the $\partial^\nu \partial_\mu \omega^\mu$ term in the equations of motion,
which should not change the time-evolution of our fields, provided the gauge condition
is always verified. This is in principle the case if the initial condition verifies
the gauge condition. Numerically, the system tends to drift away from this condition
and one needs to take additional precautions to enforce the gauge condition at all times.

To enhance stability, we introduce the ``electric'' and ``magnetic'' fields of $\omega$, 
${\cal E}_k = \partial_k \omega_0 - \dot{\omega_k}$, ${\cal M}_k = \epsilon_{klm} \partial_l \omega_m$,
and eliminate $\omega_0$ as a dynamical variable. The latter is possible because the
mass enforces the gauge $\dot{\omega_0}=\partial_k \omega_k$ effectively leaving only three
dynamical degrees of freedom for the $\omega$ field. 
This choice of variables allows us to use a local scheme, which gives the
new time derivatives at a given spatial point as an implicit function of the 
local time derivatives and spatial derivatives only.

The $\Psi$ equations of motion are in a form which ensures the chiral condition without having 
an explicit Lagrange multiplier. We discretize the fields on a uniform three-dimensional 
grid. The values of all fields at a given timestep are defined on each grid point. 
The time-derivatives of the fields are retarded with one half timestep. The time-evolution
of the fields themselves is thus straightforward. The velocities are evolved using the 
second-order equations of motion. This involves solving locally a set of coupled implicit
equations, since the velocities also appear in the interaction terms. 

In the $\Psi$ scheme, our set of equations is
\ber
\ddot{\Psi}^A ~&=&~ \sum\limits_k \partial_k^2 \Psi^A 
- \Psi^A \sum\limits_E \lp \dot{\Psi^E}^2 - \sum\limits_k \partial_k {\Psi^E}^2 \rp
+ \mpi^2 \lp \delta^{A0} - \Psi^A \Psi^0 \rp \nonumber\\
&~&~~~+ \frac{3 g}{8 \pi^2 \fpi^2}
\epsilon^{ABCD} \Psi^B \lp \epsilon_{klm} {\cal E}_k \partial_l \Psi^C \partial_m \Psi^D
+ 2 {\cal M}_k \dot{\Psi}^C \partial_k \Psi^D \rp 
\nonumber \\
\dot{{\cal E}_k}~&=&~ - \epsilon_{klm} \partial_l {\cal M}_m - M^2 \omega_k
+ \frac{ 3 g}{8 \pi^2} \epsilon_{klm} \epsilon^{ABCD} \Psi^A \dot{\Psi}^B \partial_l \Psi^C 
\partial_m \Psi^D
\nonumber \\
\dot{{\cal M}_k}~&=&~\epsilon_{klm} \partial_l {\cal E}_m \nonumber \\
\ddot{\omega}_k~&=&~ \dot{{\cal E}_k} + \partial_k \partial_l \omega_k~~.
\eer
In addition, we compute $\omega_0$ using the gauge condition and use it in the evaluation
of the energy.

Both in the choice of the algorithm and of the
process we tried to keep things conceptually as simple a possible, in order to
avoid additional sources of perturbation and error.
Thus we did not take advantage of the cylindrical symmetry of the head-on process,
even though most of our calculations have been done for this case.
The cylindrical coordinate system has been cited as a possible source of instability 
\cite{Livermore}. 

The technical setup of the calculations was the same as described in \cite{SkSk}. 
We used clusters of typically six to ten IBM SP-2 machines.
For the final run we used a cluster of 20 machines.
Our parallel codes are written in Fortran 90. We use a variety of grid sizes. 
Our physical box is $10 \times 10 \times 10$ fermi for the calculation itself, but
for the scaling analysis presented in Subsection 3.3 it is
significantly smaller, only $6 \times 5 \times 5$ fermi. 
By taking advantage of the symmetry of the problem,
we actually simulate only an eighth of the physical box. We used various
grid sizes, from 8 to 20 points per fermi, 12 points/fermi for the main runs.
We used timesteps varying from $100$ per fermi to $3200$ per fermi in the violent
regime.
As initial conditions we use a static, spherically symmetric soliton whose profile
we obtained in a separate calculation. We boost this configuration to $\beta = 0.5$
towards its mirror image created by the appropriate boundary conditions at the 
symmetry wall. 
The total initial energy of the two soliton system is then $1500 \mev$, of which $1300 \mev$
is in the solitons and approximately $200 \mev$ is kinetic energy.

\subsection{Results}

In this paper we refer only to head-on collision, even though our codes
allow for any initial conditions. Below we describe the process of annihilation of 
a Skyrmion and an anti-Skyrmion which are initially $3.0 \fm$ apart and are boosted
towards each other with an initial velocity of $\beta=0.5$ each. 
\begin{figure}
\centerline{
\vbox{
\hbox{
\psfig{file=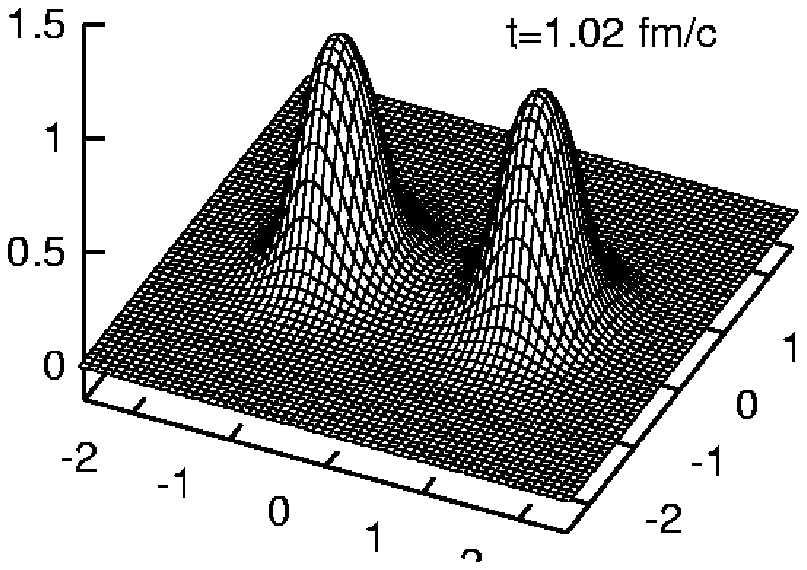,width=5.5cm,angle=0}
\psfig{file=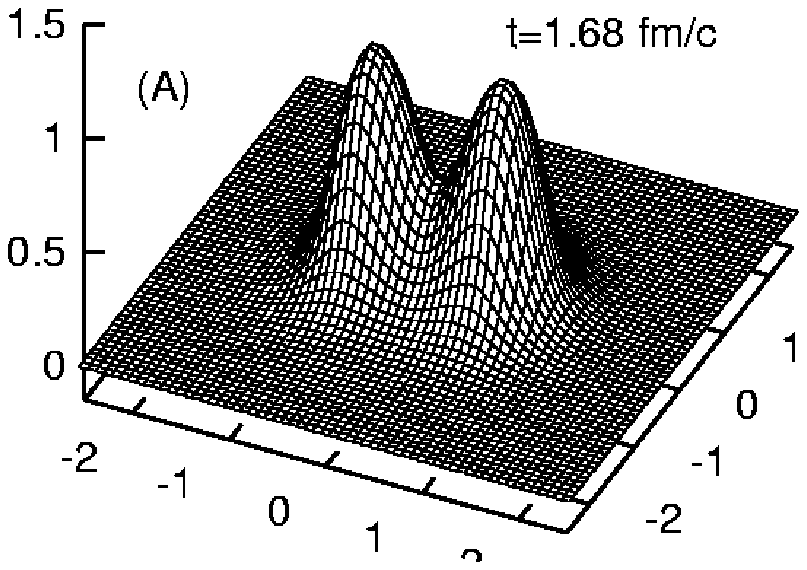,width=5.5cm,angle=0}
\psfig{file=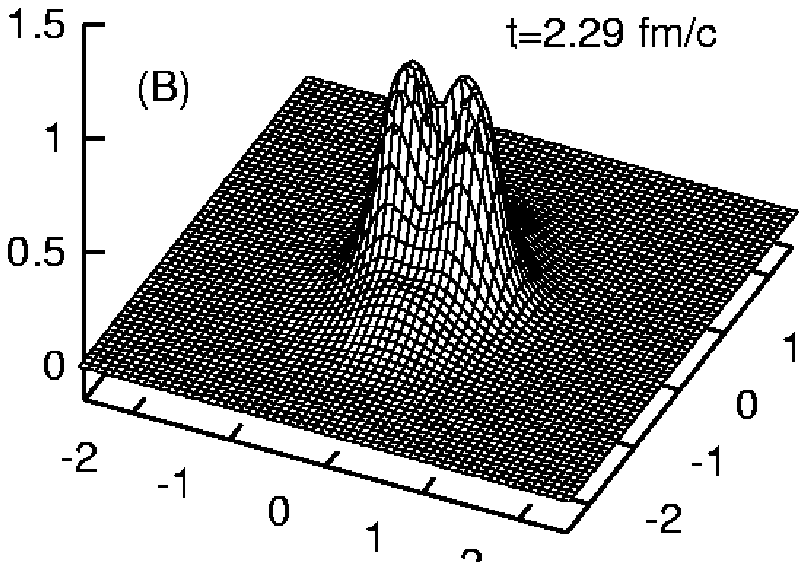,width=5.5cm,angle=0}
}
\hbox{
\psfig{file=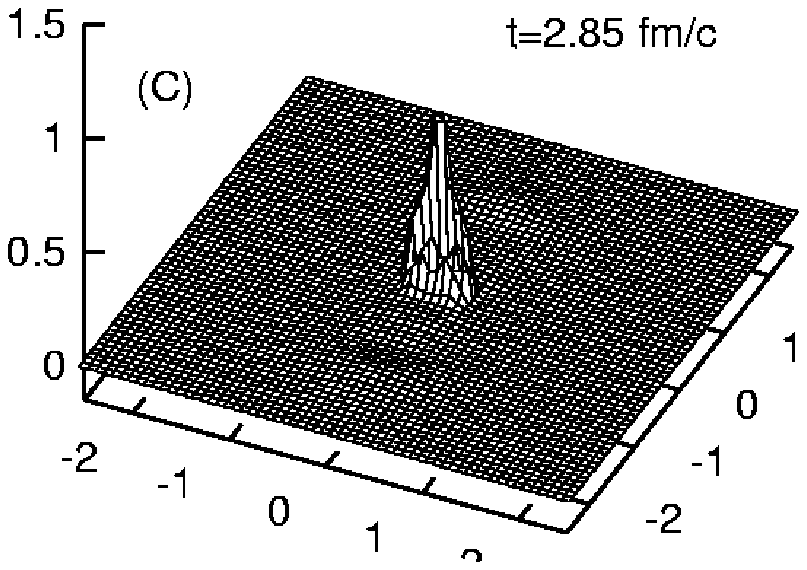,width=5.5cm,angle=0}
\psfig{file=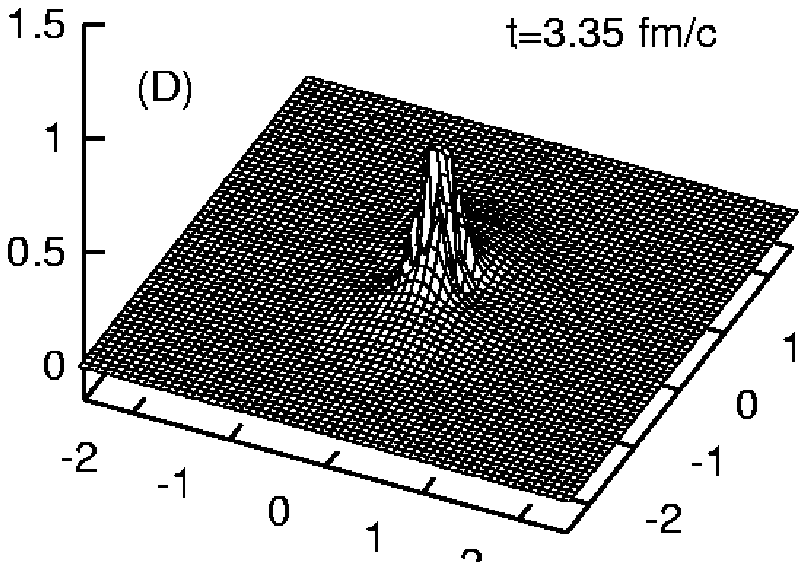,width=5.5cm,angle=0}
\psfig{file=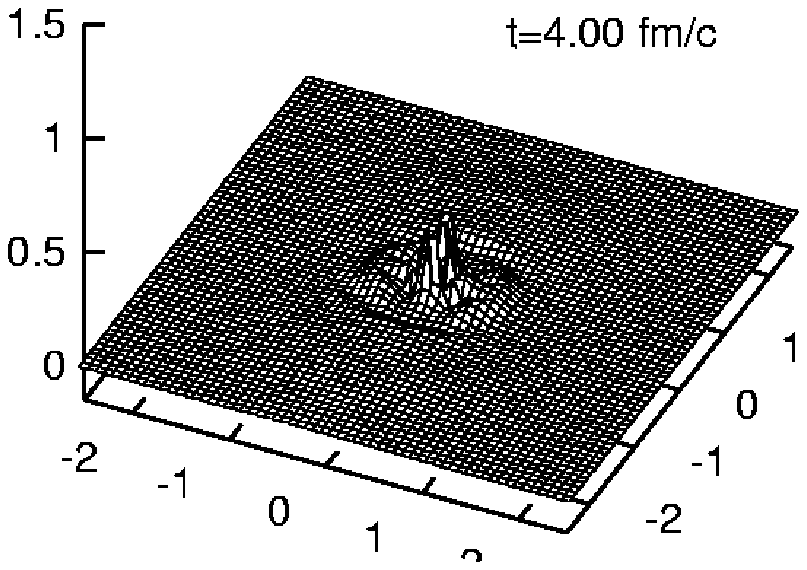,width=5.5cm,angle=0}
}
\hbox{
\psfig{file=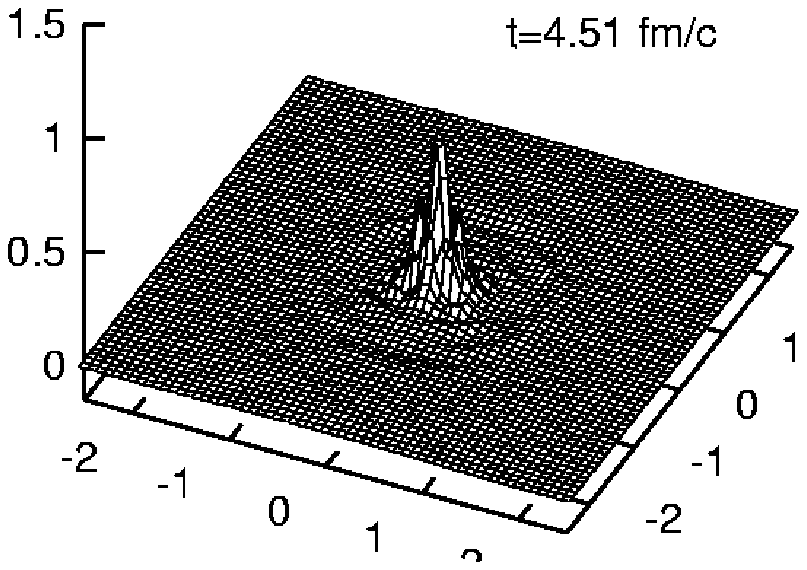,width=5.5cm,angle=0}
\psfig{file=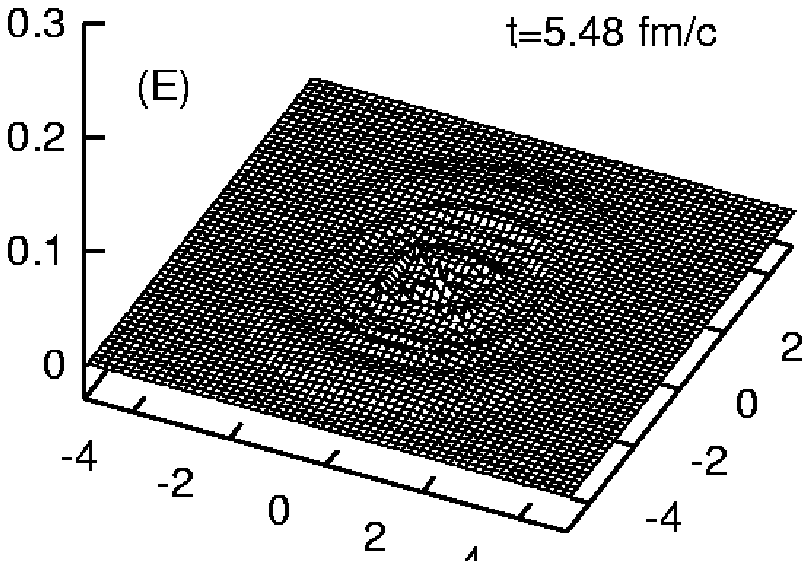,width=5.5cm,angle=0}
\psfig{file=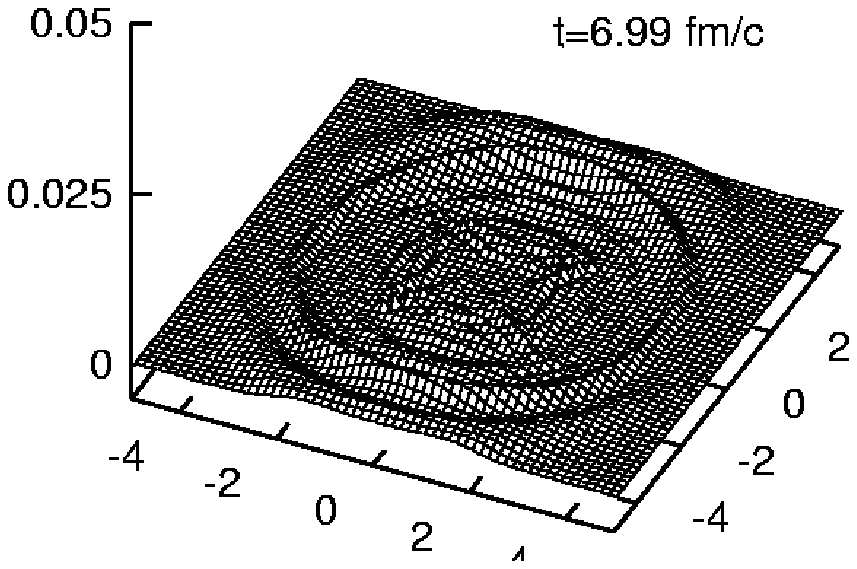,width=5.5cm,angle=0}
}
}
}
\caption{
One component of the pion field  $(1 - \Psi_0)$ at various moments during annihilation.
The fields are shown in the $xy$ plane. The $x$ axis is the direction closer 
to horizontal. The length on the axes is measured in fermi. The quantity we plot is dimensionless. 
Note the different vertical and horizontal scales in the last two frames.
}
\label{Phi3d}
\end{figure}

The process is best illustrated by the time 
evolution of the pion field. 
In Figure \ref{Phi3d} we plot the
quantity $(1-\Psi_0)$ as a function of time.
This choice is natural 
 since in free space $\Psi_0=1$ and at the
tip of a Skyrmion $\Psi_0=-1$. 
In the first plot we have the essentially unmodified
Skyrmion and anti-Skyrmion, slightly superimposed. The $\Psi_0$ component of the pion
field is identical for the two objects. 
Charge conjugation and grooming affects only
the 'spatial' components. 

In order to annihilate, the fields have to unwind\footnote{
Because of symmetry, on the central axis the field is always of the form 
$\Psi=(cos\theta,sin\theta,0,0)$ . As one passes through the center of a Skyrmion, the angle rotates
through a full circle. For the anti-Skyrmion, the winding is opposite. As the two objects
approach, the center point unwinds. 
}, therefore the
field in the center point has to pass through the value $\Psi=(-1,0,0,0)$ (the highest point
in our plots). In other words, the tips must merge before unwinding.
The second frame illustrates a moment close to this situation.
                     As    will be discussed
later, the axial dependence is rather sharp at this moment. In our simulation the symmetry
center is between lattice points, therefore the crest in the second frame should be close
to horizontal with a proper extrapolation. 

\begin{figure}
\centerline{
\vbox{
\hbox{
\psfig{file=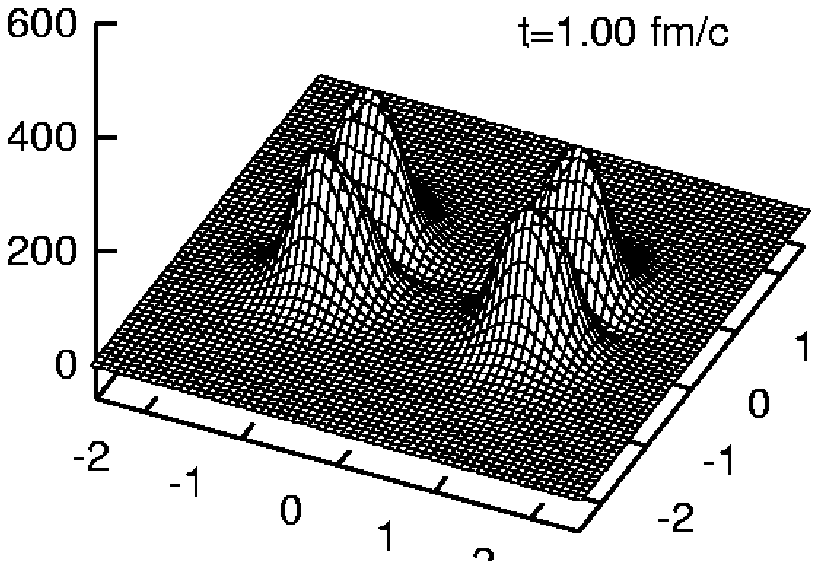,width=5.5cm,angle=0}
\psfig{file=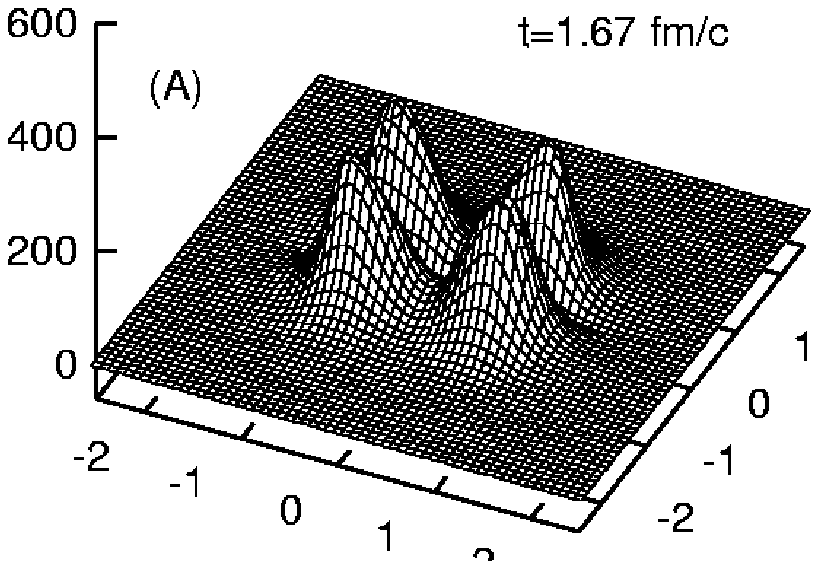,width=5.5cm,angle=0}
\psfig{file=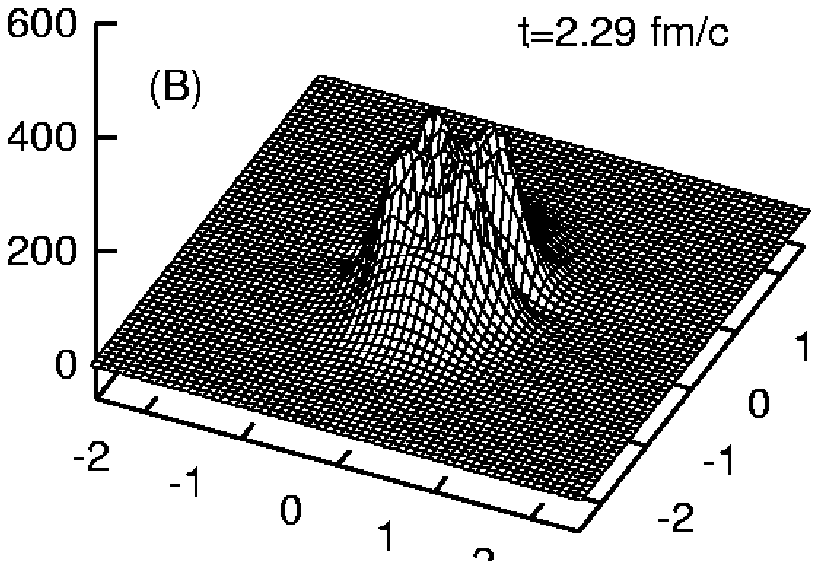,width=5.5cm,angle=0}
}
\hbox{
\psfig{file=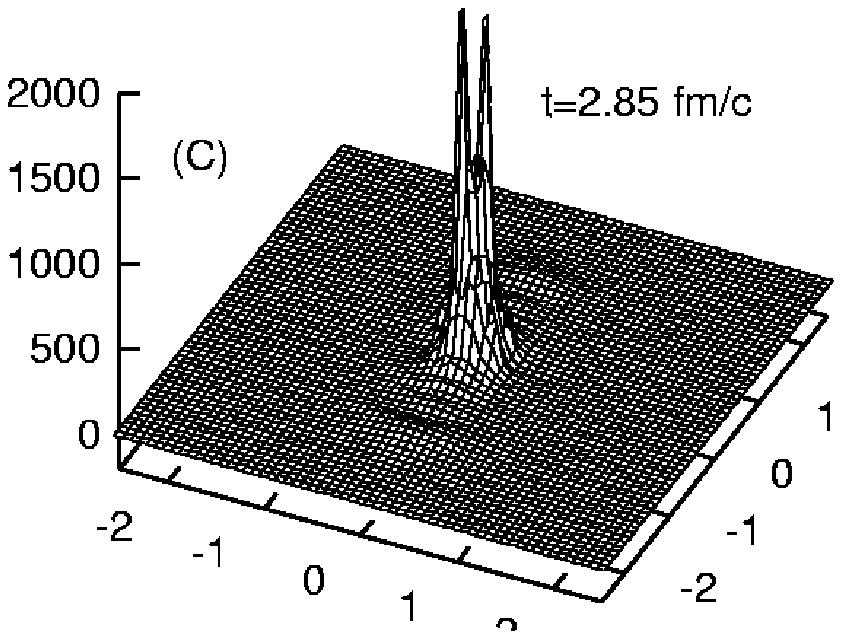,width=5.5cm,angle=0}
\psfig{file=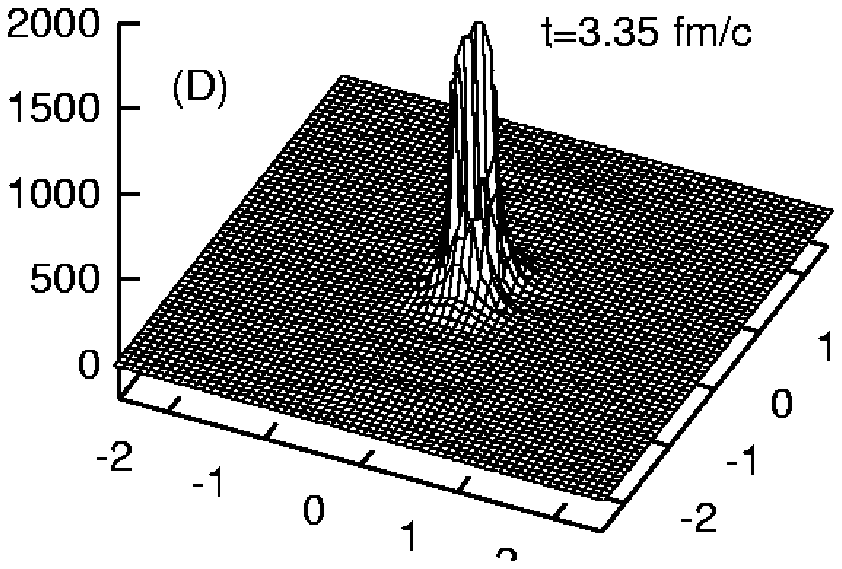,width=5.5cm,angle=0}
\psfig{file=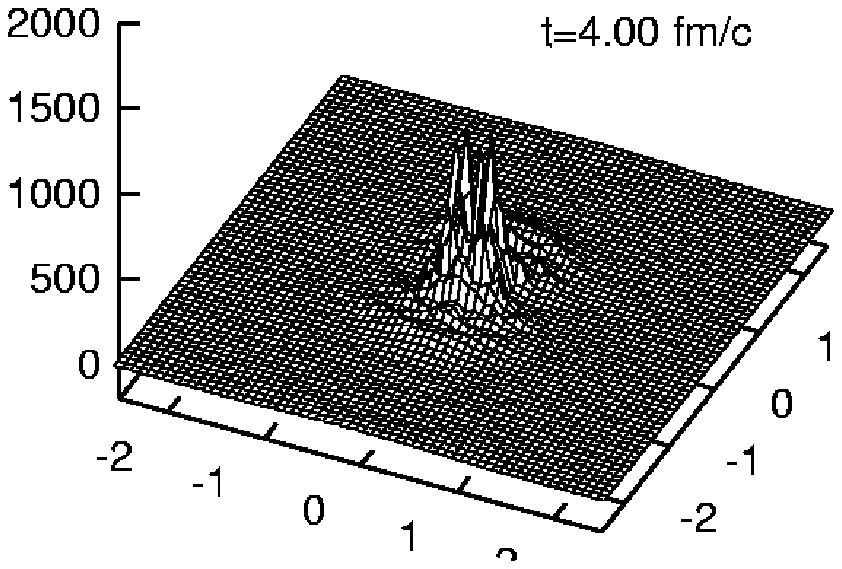,width=5.5cm,angle=0}
}
\hbox{
\psfig{file=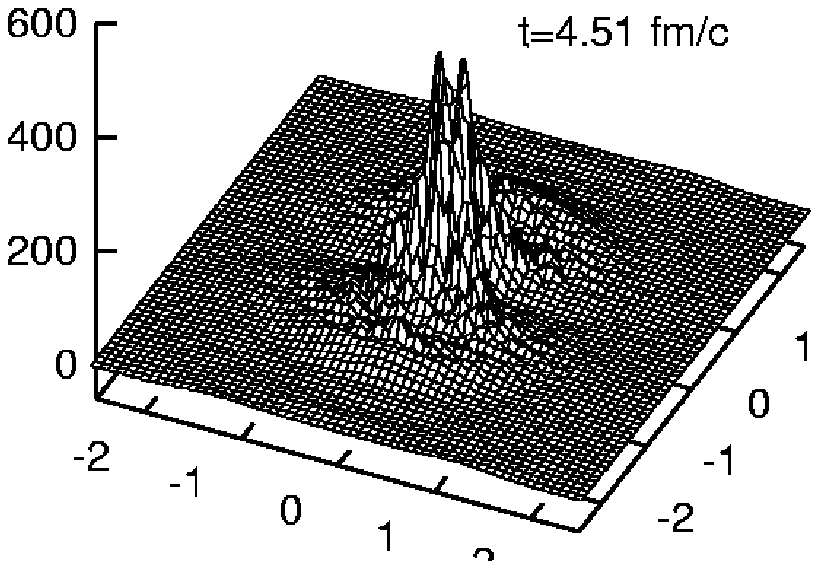,width=5.5cm,angle=0}
\psfig{file=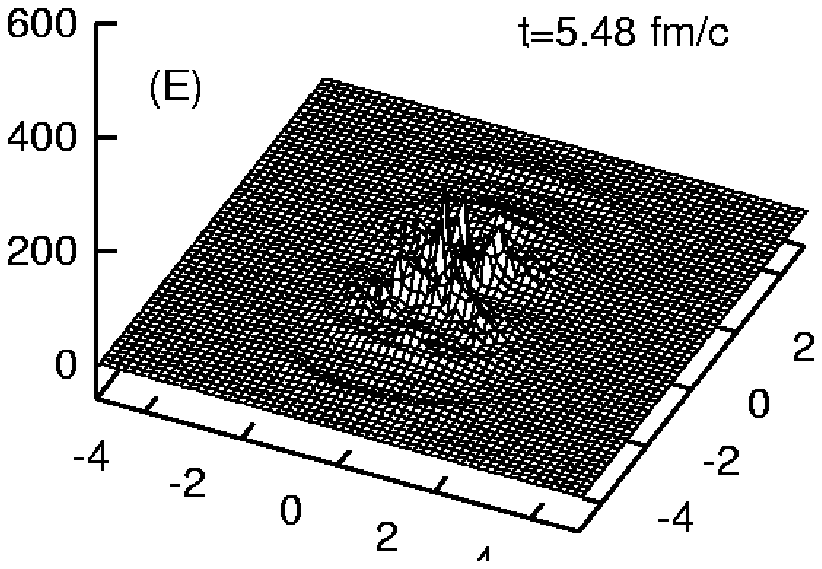,width=5.5cm,angle=0}
\psfig{file=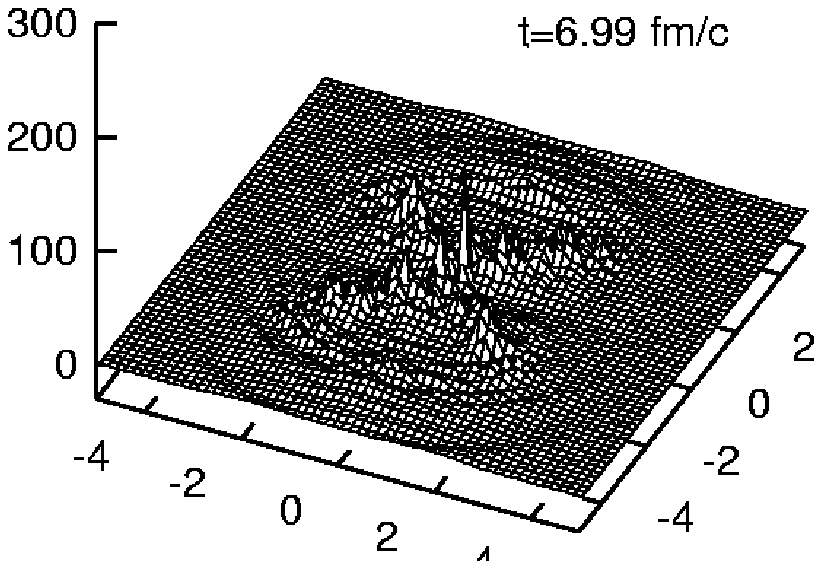,width=5.5cm,angle=0}
}
}
}
\caption{
The total energy density multiplied by the distance to the symmetry axis,
in $\mev / \fm^2$ 
at various moments during annihilation. 
The density is shown in the $xy$ plane. The $x$ axis is the direction closer 
to horizontal. The length on the axes is measured in fermi. 
Note the different vertical and horizontal scales in the last two frames.
}
\label{E3d}
\end{figure}

From this point on, the value of $\Psi_0$ approaches
fast the vacuum value as the topological obstacle is now gone.
After another fm/c the field is close to the vacuum value $(1,0,0,0)$. 
The variation is so fast that the field in the center 'overturns' and increases again. 
Only after 3 fm/c from the passing of the peak do these large amplitude oscillations 
subside by propagating outwards as spherical waves.
This outgoing pion radiation is clearly seen in the final two frames of Figures \ref{Phi3d} 
and \ref{E3d}.

\begin{figure}
\begin{center}
\psfig{file=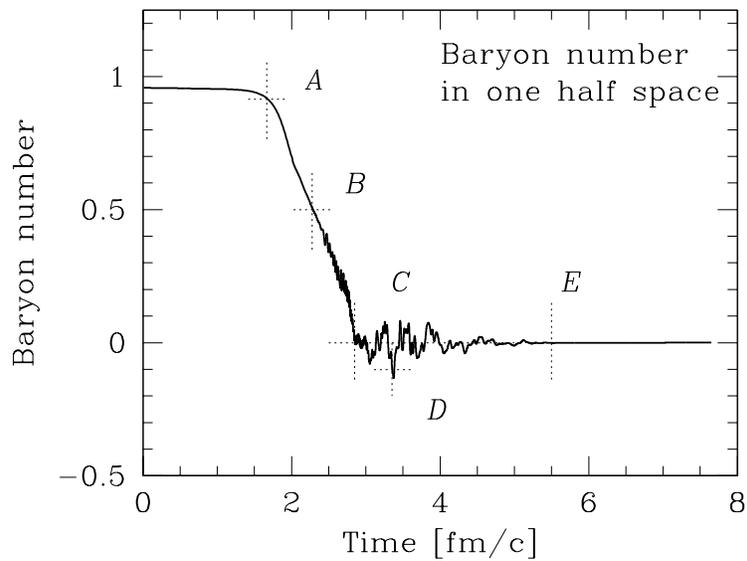,width=4.0in}
\caption{
Evolution of the baryon number in one half-space during a central annihilation process.
The points $A \ldots E$ indicate particular moments which are also indicated in Figure \ref{Phi3d},
\ref{Evst}, and so on, 
and are referred to throughout the text. Note especially the point $B$, which corresponds to
'half-annihilation'. It is associated with the merging of two topological centers and also
marks the beginning of the violent part of the annihilation process.
}
\label{Bvst}
\end{center}
\end{figure}

The time
evolution of the baryon number is illustrated in Fig.\ref{Bvst}.
We compute the baryon number by integrating (\ref{barydens}) in one half-space.
The baryon number in the other half space is equal and opposite to this.
The annihilation starts basically when the Skyrmions touch, at $t=1.66 \fm$ 
(point $A$ in Fig.\ref{Bvst}).
In the absence of interactions, exactly half of the baryon number in one half
space should annihilate when the two centers coincide, at $t=3.0 \fmc$ . 
Because of the attractive interaction, this happens a little faster, at around 
$t=2.3 \fmc$ ($B$). Along with the field, the remaining baryon number decreases 
quickly to zero ($t=2.85 \fmc$, point $C$), continues to decrease for a 
short time, then oscillates, hitting an absolute minimum at point ($D$). 
The baryon number oscillates
along with the large amplitude oscillations of the field and finally settles at zero
($t=5.5 \fmc$, point $E$) in the radiation regime. 

Altogether, the unwinding of the field (from $A$ to $C$) takes approximately 
$1.2\ldots1.3 \fmc$, but this is followed by localized oscillations which take a
longer amount of time ($C$ to $E$, approximately $2.5 \fmc$), therefore the total
process from the moment when the Skyrmions touch to the complete disappearence of 
the baryon number takes about $3.5 \fmc$, depending on the choice of the point $E$.

\begin{figure}
\begin{center}
\psfig{file=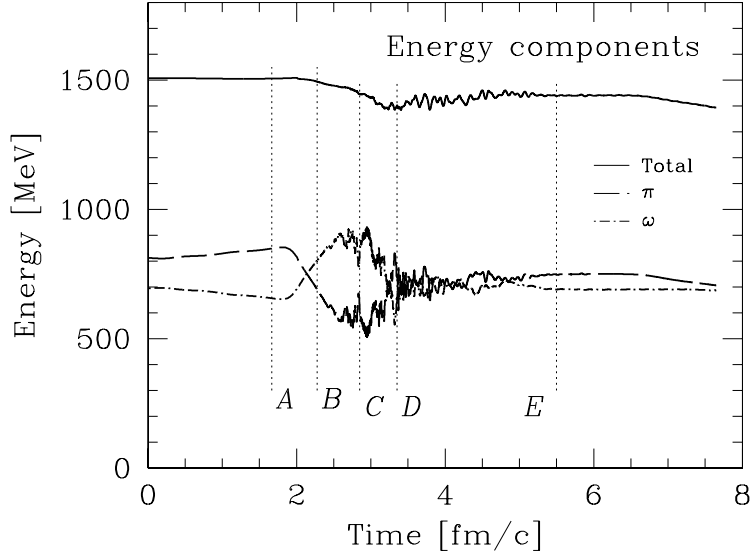,width=4.0in}
\caption{
Evolution of the energy and its components
during a central annihilation process.
}
\label{Evst}
\end{center}
\end{figure}

In Fig.\ref{Evst} we plot the evolution of the energy as the sum of the energy in the pion field
and the omega field. The dotted lines and labels $A\ldots E$ on 
Fig.\ref{Evst} indicate the same times as in Fig.\ref{Bvst}. 
Our definitions of the energy densities 
-- which are integrated numerically to give the quantities in Figure \ref{Evst} --
are
\ber
{\cal H}_\pi~&=&~\frac12 \lp \dot\Psi^A \dot\Psi^A 
+ \sum\limits_{k=1}^3 \partial_k \Psi^A \partial_k \Psi^A \rp \nonumber \\
{\cal H}_\omega~&=&~\frac12 \sum\limits_{k=1}^3
\lp M_\omega^2 \omega_k^2 + {\cal M}_k^2 + {\cal E}_k^2 \rp 
+ \frac12 M_\omega^2 \omega_0^2 ~~.
\eer
The piece corresponding to the $\omega_0$ field can be defined in terms of our dynamical variables,
\ber
{\cal H}_{\omega_0}~&=&~\frac12  M_\omega^2 \omega_0^2 = 
\frac{1}{2 M_\omega^2} \ls \lp \frac{3 g }{2} B_0 \rp - \lp \partial_k {\cal E}_k \rp \rs^2~~.
\eer
The second identity follows from the equations of motion
for $\omega_0$, the definition of ${\cal E}_k$, and the gauge condition.\footnote{
Numerically this identity is violated because the right-hand side is the difference
of two large numbers in the fast-varying regime. 
The identity relies on third-order derivatives of our dynaimcal quantities.
However, using $\omega_0$ computed from the
gauge condition leads to reasonable energy conservation.}
We also plot the total energy in Fig.\ref{Evst}. 
There is a loss of less then $100 \mev$ from a total
of $1500 \mev$ between the points $B$ and $E$, which corresponds to 
approximately $7 \%$\footnote{The loss in the run presented in this section is less, actually closer to
$5 \%$, but the runs presented in the next section, which use a smaller physical box and slightly
different initial conditions, lose about $100 \mev$.}. We
assign this loss to numerical dissipation which is significant in the fast-varying regime between
half-annihilation ($B$) and the onset of the radiation regime ($E$). 
The further decay of the total energy simply corresponds
to outgoing radiation which leaves the simulation box.

The annihilation process is accompanied by the rearrangement of the energy between the pion
and omega sectors. Initially we have free-space propagation 
of the solitons. 
The smooth part of the unwinding beginning at ($A$) is accompanied by the flow of energy 
from the pion field into the omega sector. The net flow stops before full unwinding
($C$) and yields to oscillations which correspond in time to the large oscillations of the
field, accompanied by significant oscillations of the baryon number around zero.
During this regime, when the net baryon number oscillates,
the energy also flows back and forth between the two sectors.
Eventually the two sectors stabilize after $E$ at comparable values.

A more detailed picture of the energy flow is given by looking at the spatial distribution of the
energy at various moments. In Figure \ref{E3d} we plot the total energy
density  multiplied with
the distance $\rho$ to the symmetry axis. Plotting $\rho \times dE / dV$  gives a better 
estimate of the relative amount of energy contained in different regions of space. 
In the first three frames we see the two configurations approaching each other, then 
merging. The void in the middle of each lump is due to the fact that we multiplied the energy
density with the distance to the $x$ axis. In spite of this, later on the strength is
concentrated in the center. Eventually the energy starts to flow outwards in almost spherical waves.

These plots reveal two important aspects. First, the fact that the energy 
density is very high
in the center between $t=3.0 \fmc$ and $t=5.0 \fmc$, which is the
period between annihilation and the start of
significant outward energy flow. 
Note the higher enegy scale on the middle three plots.
The other important feature is the abundance of fast, small
amplitude oscillations which persist to the end of the time interval under consideration.
They can be seen well in the last three plots.
These oscillations originate in the period immediately following annihilation when there 
is a very fast,
global variation of the fields confined to a small region of space.
Most likely numerical error stemming from the large local variations has a role as a source
for these oscillations. However, as discussed in the next subsection on 
numerical stability, the persistence of the small oscillations is also possibly  due to 
properties of the exact, continuum equations of motion.

Despite the presence of the small oscillations, there is a well defined pattern to the 
flow of energy, both in terms of radial flow and angular distribution. Starting with $t=2.5 \fmc$,
the energy is concentrated in a small region (initially of radius $1 \fm$) around the center.
The distribution becomes very strongly peaked in the time between half-annihilation and complete
unwinding. There is some outgoing radiation as early as $t=3.5 \fmc$ but the strongly peaked 
pattern eases up only after $t=5.0 \fmc$ when the bulk of the energy starts flowing outward.
\begin{figure}
\begin{center}
\psfig{file=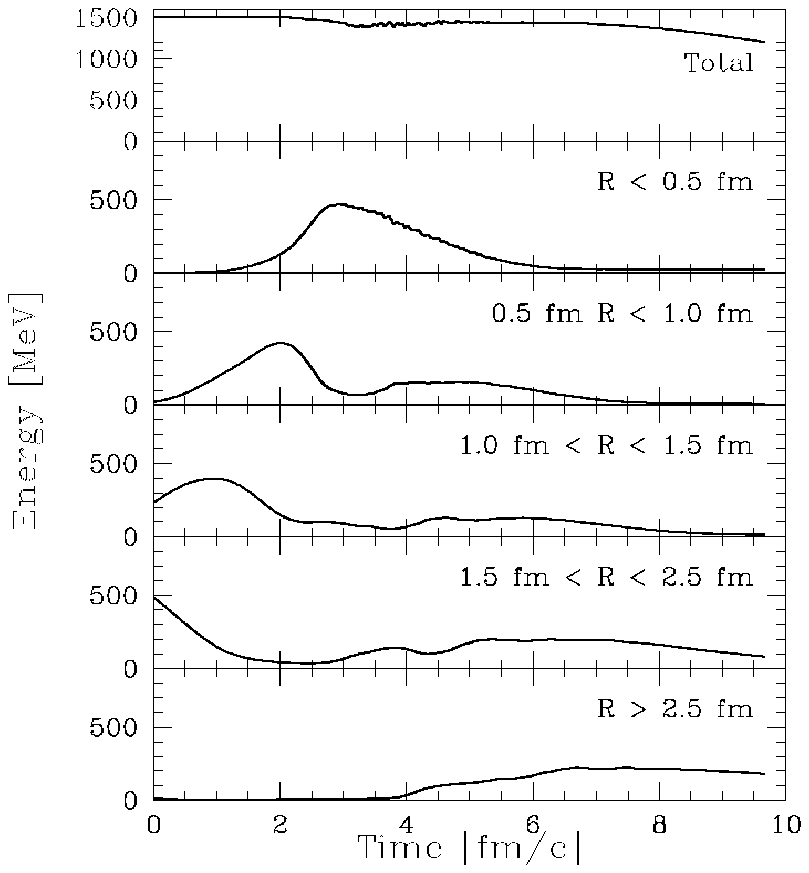,width=5.0in}
\caption{
Total energy in spherical shells surrounding the annihilation as a function of time.
}
\label{shells}
\end{center}
\end{figure}
The macroscopic flow of energy is nicely illustrated in Fig.\ref{shells}  by plotting the 
total energy contained in spherical shells surrounding the center. 
Initially we have the two incoming solitons which move through bins in decreasing order
of the radius. Starting from $t=2.5 \fmc$ the energy accumulates in the center bin 
(a sphere of radius $0.5 \fmc$, half the linear size of a Skyrmion). The accumulation peaks
at about $t=3 \fmc$ when this bin contains more than $2/3$ of the total energy. 
Even though the outward flow starts as early as $t=3.0 \fmc $, it becomes significant only later. 
Approximately $90 \%$ of the energy leaves the center sphere by $t=6.0 \fmc $ , or 
$3.0 \fm$ after full unwinding.
From then on, we can follow the energy moving outwards through bins
of increasing radius. We can also understand how the decrease in total energy (top panel, same
line as in Fig.\ref{Evst}) is really just flow leaving the simulation box.

\begin{figure}
\begin{center}
\psfig{file=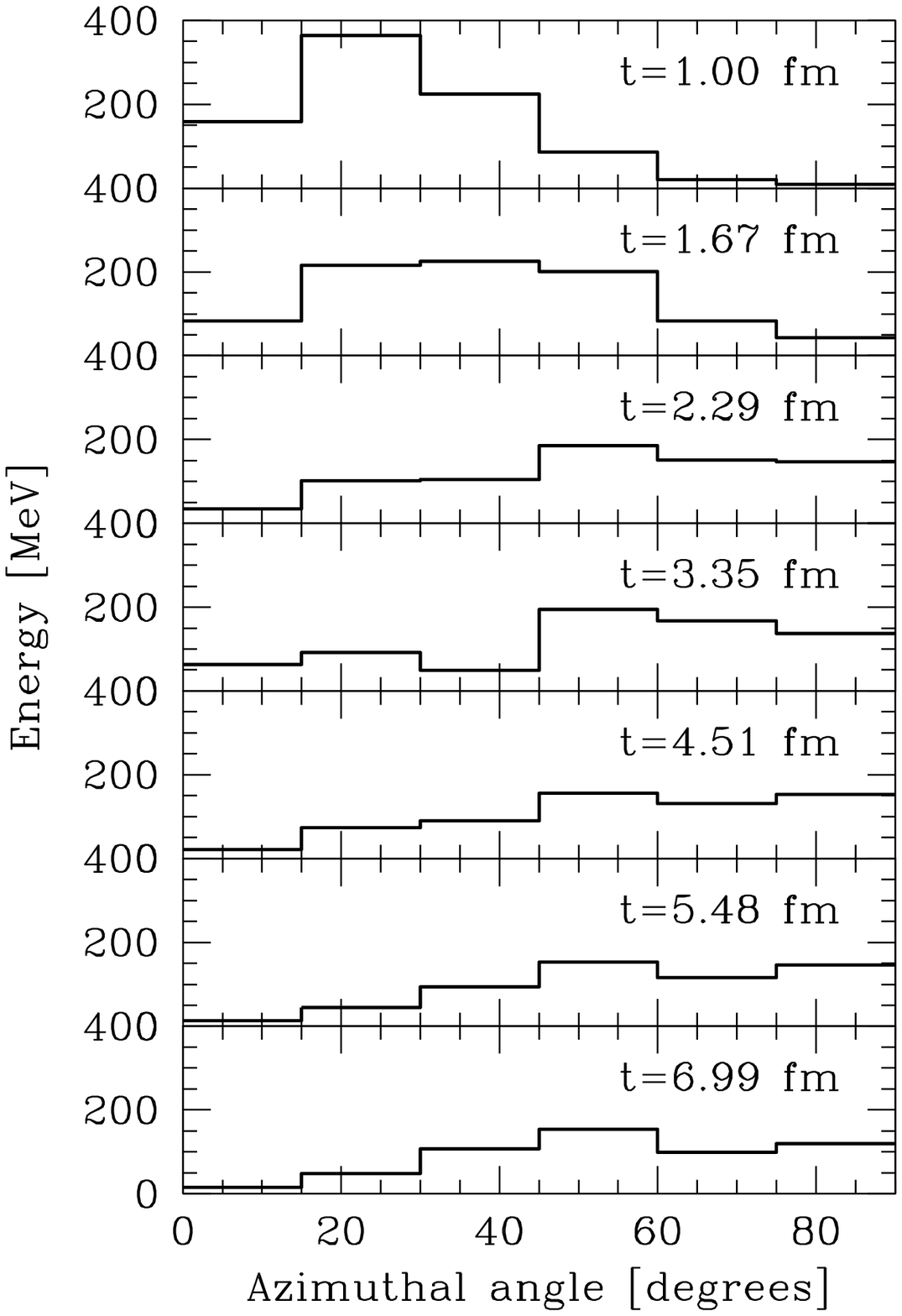,width=5.0in}
\caption{
Total energy in angular intervals measured from the symmetry axis at various moments during
annihilation.
}
\label{angleshisto}
\end{center}
\end{figure}

\begin{figure}
\begin{center}
\psfig{file=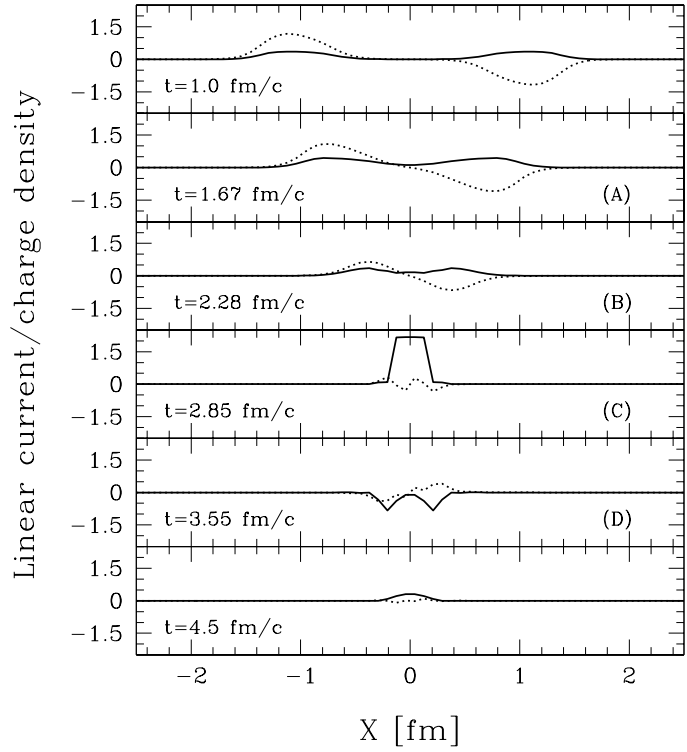,width=4.0in}
\caption{
The baryon number density (dotted) 
and the $x$ component of the baryon current density (solid), integrated over $y$ and $z$
(in $\fm^{-1}$ respectively ${\rm c} / \fm$),
as a function of the $x$ coordinate, at various moments during the annihilation process.
}
\label{angles}
\end{center}
\end{figure}
Finally, in Figure \ref{angleshisto}  we illustrate the angular distribution 
of the energy during the annihilation process. In Fig.\ref{angleshisto} we can see
a depletion of the small angle bins, and a peak around $45^o$ which seems to be the preferred
direction of the energy flow.  It should be noted
that these angular distributions are in a plane, but that the three dimensional 
distribution has cylindrical symmetry about the direction of collision.

\subsection{Checks}

In this subsection we investigate the extent to which our results are influenced
by the details of the numerical calculation. This aspect is particularly important
because of the presence of large time and spatial derivatives of the fields. 
The macroscopic dynamics of the problem impose a certain mimimum size for the
simulation box (a cube of size $5 \fm$). We use a fixed grid. This limits the number
of points per fermi we can have to not much more than 20. Our typical calculations
use 12 points per fermi. 

One obvious concern stems from the fact that we were able 
to ensure energy conservation
only to about $7 \% $, at best $5 \%$. 
Approximately $70\ldots 100 \mev$ of the total initial energy of $1500 \mev$
is lost to numerical dissipation, as  can be seen in Figs.\ref{Evst} and \ref{shells}. This loss
comes between
the points $B$ and $C$ in the respective plots, and is a significant but not
alarming energy loss. 

Numerical error is probably also responsible for the appearence of persistent 
but random oscillations.
Starting with $t=4.0 \fmc$, the field configurations display oscillations on the scale of
a few lattice spacings. This is preceded by a configuration which is probably singular
in the continuum limit, at half-annihilation (we discuss this in a separate section).
This raises the question of whether the continuum physics we wish to study mixes with 
lattice artifacts, or rather, whether we are at able to extract continuum physics reliably.

To test the stability of our results, 
we performed several runs using the $\pi$ algorithm for a number of different lattice
spacings. 
The results obtained with this algorithm with identical box size and initial conditions 
are very similar to those obtained with the $\Psi$ algorithm. 
The program based on the $\pi$ algorithm has been readily available for a longer time, 
and we used it
for the very time consuming runs with larger numbers of points.
Below we present results for 12, 16 and 20 points per fermi. The physical 
process in these runs is the same, i.e., head-on annihilation with the same parameters as 
in the main run, initial separation of $3 \fm$, initial velocity of $\beta=0.5$. 
However, we used the smallest possible simulation box, only $6 \times 5 \times 5 \fm^3$,
and a slightly different initial configuration.
Using an even smaller box would have resulted in more energy dissipation outside the box,
and most notably, reflection off the walls (which we are unable to eliminate completely)
which would interfere with the 'real' physics. We then compared the results from these
various runs. Ideally,
the numerical artifacts should scale away as the lattice spacing is decreased. 
\begin{figure}
\begin{center}
\psfig{file=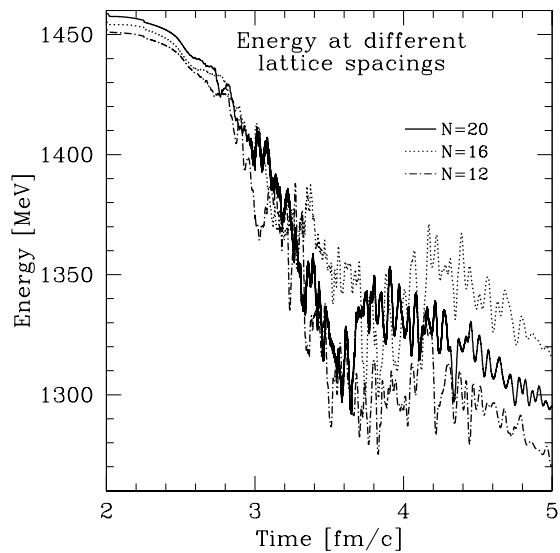,width=3.0in}
\caption{
The total energy during the more violent part of the annihilation process,
in three calculations which are identical except for the number of points
per fermi, $N$. Note that these runs are slightly different from the runs discussed 
in the preceding section. They are performed in a smaller box using a slightly different
initial configuration.
}
\label{energyscale}
\end{center}
\end{figure}
In Figure \ref{energyscale} we plot the total energy for three runs which are identical
(including the timestep, which is decreased before $t=3.0 \fmc$ from $100$ per fermi to 
$3200$ per fermi/c) except for the lattice spacing which is respectively $N=12, 16, 20$ points
per fermi. We zoomed in to the region between $t=2.0 \fmc$ and $t=5.0 \fmc$, when most of
the dissipation takes place. The energy decreases as the system is squeezed into the small
region around the center. The loss is likely to be caused by discretization error. 
However, as the system
expands, only a fraction of the loss is recovered. 
The same run with $N=8$ points per fermi, which we cannot plot 
 on the same graph without making the graph completely unintelligible, gives 
worse conservation than all three runs plotted here. The two larger spacings do give better
energy conservation then the $N=12$ run, as one would expect. 
However, there is no clear scaling, as the order of the $N=16$ and $N=20$ results
is reversed. As an argument for the reliability of our calculations, let us emphasize that
we are talking about differences of $30 \mev$ here between 
calculations, i.e., $2 \%$ of the
total energy.\footnote{We remind the reader that these plots refer to a calculation similar to the one described in the previous subsection but in a smaller simulation box. Therefore, these runs also have some additional energy flowing out from the simulation box, even though most of the loss 
in this period is through numerical dissipation.}
\begin{figure}
\begin{center}
\psfig{file=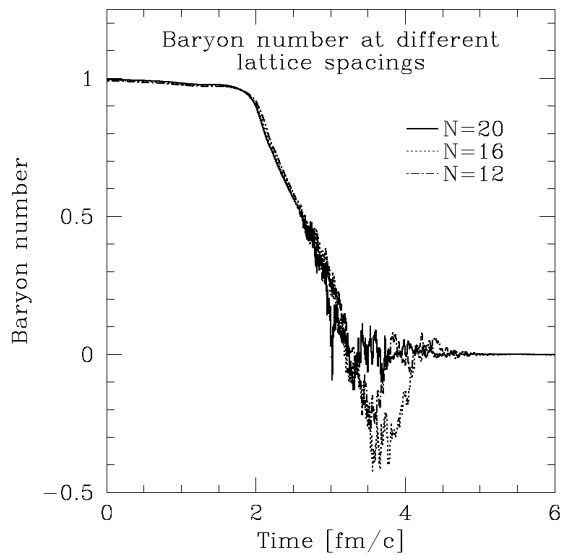,width=3.0in}
\caption{
The baryon number in one half space during the annihilation process,
in three calculations which are identical except for the number of points
per fermi, $N$.
 Note that these runs are slightly different from the runs discussed 
in the preceding section. They are performed in a smaller box using a slightly different
initial configuration.
}
\label{barynoscale}
\end{center}
\end{figure}
A look at the comparative plot of the baryon number in the same runs reveals a 
similar picture. In all three runs shown, the remarkable points $A$,$B$, $C$ and $E$,
i.e., the start of the unwinding, the point of half-annihilation and full unwinding, as
well as the end of the violent fluctuations, practically coincide. 
However, the extent and duration of the excursion of the baryon number below zero varies.
It is practically absent for $N=8$ but there is a large fluctuation later on (again, not plotted).
For $N=12$ and $N=16$ we see a sizeable excursion, larger for $N=16$, but again the $N=20$
calculation is out of sequence and has only a small negative excursion.
It could be that the timestep choice (same for all runs) is too large for $N=20$.   
Even with this in mind, the overall evolution of the baryon number  is very similar 
for the four runs we discussed. The decay of the baryon number and the duration of the
large oscillations is a robust feature of the calculation. Furthermore the small 
oscillations seem to be very noisy with little relationship from grid size to grid size. 
\begin{figure}
\begin{center}
\psfig{file=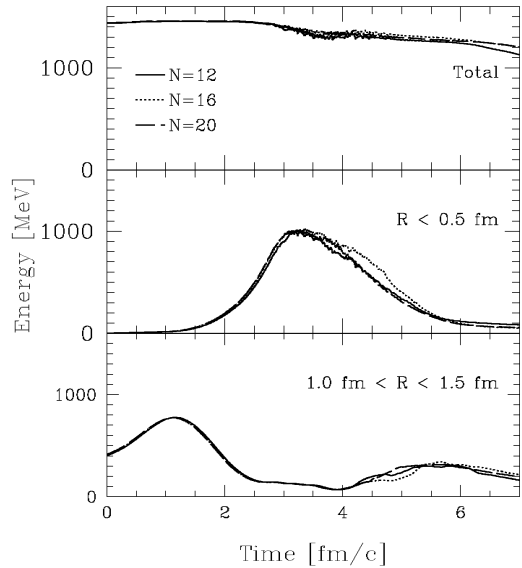,width=3.0in}
\caption{
Evolution of the energy contained in a few spherical shells for various lattice spacings.
}
\label{shellscaling}
\end{center}
\end{figure}
The outward energy flow is perhaps the most important quantitative result one may 
extract from a numerical calculation of Skyrmion-anti-Skyrmion annihilation. 
This result would be the starting point for constructing the final pion and omega
states in a calculation of low energy $N \bar{N}$ annihilation \cite{Lu&RDA}. 
In Figure \ref{shellscaling} we plot the energy contained in centered spherical shells,
for the three $N=12,16,20$ calculations discussed above. 
The evolution of the energy contained in the shells is practically the same for the three
runs. There is some fluctuation, especially for the center bin, but the differences are
not very large. In particular there is little difference
in the bins at larger distance, and that is where information would be extracted
for the outgoing pion and omega states. 
We cannot identify a scaling pattern in these cases. 
However, it is clear that the dominant process of outward flow of energy is well
established and is clearly not influenced by variation in
 the number of lattice points.

In conclusion, our detailed results are somewhat sensitive
 to the number of lattice points, but
the physically important observables, energy conservation, the time evolution of the baryon number, 
and the flow of energy are reasonably well determined and independent of variations in 
the number of lattice points.

\section{Stability analysis}

One of the two most prominent features of our results is the appearence of 
persistent, small amplitude oscillations after unwinding. These oscillations are present
practically to the end of our calculations. In the early post-unwinding regime they 
influence the total energy and baryon number.

The frequency of the fluctuations of the total energy and baryon number increases with the
number of lattice points, suggesting that these are numerical artifacts. We 
have been forced to
use extremely small timesteps in our runs (thousands of timesteps per fermi). 
However, the precautions we took did not eliminate these oscillations. On the other hand,
we are able to perform our simulations to a robust end
 even in the presence of these oscillations. Furthermore,
the macroscopic (long wavelength) aspects of the outputs are not strongly influenced by 
the number of points, suggesting that numerical artifacts do not overwhelm the continuum
physics. 

The lack of success of earlier attempts to simulate annihilation \cite{Livermore,CalTech:ann,centralsc}
has been blamed on a situation which arises in the Skyrme model \cite{Livermore}. 
Due to the nonlinear nature of the interaction term,
 the equations of motion may cease to be
of hyperbolic nature, i.e., have  second time and spatial
 derivative terms of opposite 
signs. Hyperbolic equations of motion ensure the
 existence of plane wave-like solutions (of the
form $e^{i k_\mu x^\mu}$ with real wave numer $k=\sqrt{\vec{k} \cdot \vec{k}}$) which may propagate as 
packets of quasi constant amplitude. If the sign of the second time derivative 
reverses, the wave number may become imaginary for a given wave vector $\vec{k}$, resulting
in waves with exponentially increasing amplitude. A small
 fluctuation that excites this mode
would then result in a large change in the final result. In other words,
 if the equations of
motion are not hyperbolic, the system in unstable. For a deatiled derivation of the
intability for the Skyrme model of just this sort, we refer to \cite{Livermore}. 

In the following we investigate the possibility of such an
 instabiliy occuring in our model.
Recall that we are using $\omega$ stabilization rather than a Skyrme term 
in order to avoid the damage brought on by the fourth-order interaction term.
Consider our equations of motion,
\ber
\partial_\mu \partial^\mu \Psi^A  ~-~
\Psi^A \lp \Psi^E \partial_\mu \partial^\mu \Psi^E \rp 
~&=&~
 \frac{3 g}{8 \fpi^2 \pi^2 }
\eMNAB \epsilon^{ABCD}
\partial_\nu \omega_\mu \Psi^B \partial_\alpha \Psi^C 
\partial_\beta \Psi^D 
+ m_\pi^2 \lp \delta^{A0} - \Psi^0 \Psi^A \rp 
 \nonumber \\
\partial^\nu \partial_\nu \omega^\mu 
~&=&~
 \frac{g}{8  \pi^2 }
\eMNAB \epsilon^{ABCD}
\partial_\nu \Psi^A \Psi^B \partial_\alpha \Psi^C 
\partial_\beta \Psi^D 
-  M^2 \omega^\mu 
\eer
Consider now a solution of these equations and a small, fast-varying perturbation 
added to it,
\ber
\label{fastsmall}
\Psi^A~& \rightarrow &~\Psi_0^A~+~\epsilon \phi^A ~~;~~
  |\partial_\mu \phi^A | >> |\partial_\mu \Psi^A \nonumber| \\
\omega^\mu~& \rightarrow &~\omega_0~+~\epsilon \sigma^\mu~~;~~
| \partial_\mu \sigma_\nu | >> | \partial_\mu \omega_\nu |
\eer
where $\epsilon$ is a small real number.
We study the stability of the equations of motion by analyzing the behavior of
these small perturbations in the background field given by 
$\lb \Psi^A, \omega^\nu \rb $. First, we substitute the ansatz (\ref{fastsmall})
into the equations of motion and expand to first order in $\epsilon$,
\ber
\label{firstorder}
\partial_\mu \partial^\mu \phi^A ~&+&~
\phi^A \lp \partial_\mu \Psi^E \partial^\mu \Psi^E \rp ~+~
2 \Psi^A \lp \partial_\mu \phi^E \partial^\mu \Psi^E \rp~=~
\nonumber \\
~&=&~\frac{3 g }{8 \pi^2 \fpi^2} \eMNAB \epsilon^{ABCD} 
\lb \partial_\nu \sigma_\mu \Psi^B \partial_\alpha \Psi^C 
\partial_\beta \Psi^D ~+~ 
\partial_\nu \omega_\mu 
\lp \phi^B \partial_\alpha \Psi^C \partial_\beta \Psi^D
~+~
2   \Psi^B \partial_\alpha \phi^C \partial_\beta \Psi^D \rp \rb
\nonumber \\
&~&~~~~ - 2 m_\pi^2 \lp \phi^A \Psi^0 + \Psi^A \phi^0 \rp~~
\nonumber \\
\partial_\nu \partial^\nu \sigma^\mu ~&=&~
\frac{ 3 g }{8 \pi^2} \eMNAB \epsilon^{ABCD}
\lp \partial_\nu \phi^A \Psi^B + \frac13 \partial^\nu \Psi^A \phi^B \rp
\partial_\alpha \Psi^C \partial_\beta \Psi^D~~~.
\eer
We must again remember that the variation of $\Psi$ is constrained,
\ber
 ( \Psi^A + \epsilon \phi^A)  ( \Psi_A + \epsilon \phi_A) 
~=~ 1 ~+~ {\cal O} (\epsilon^2) ~~\rightarrow~~ \phi^A \Psi^A ~=~0~~.
\eer
Therefore  the quantity 
$ \epsilon^{ABCD} \partial_\nu \Psi^A \phi^B \partial_\alpha \Psi^C 
\partial_\beta \Psi^D $ has to vanish since it is the quadruple product of 
four isospin vectors which are all perpendicular ot ${\bf \Psi}$. 

We now assume that small perturbations are well approximated by plane waves,
\ber
\label{planewave}
\phi^A~=~\Phi^A e^{i k_\mu x^\mu }~~;~~
\sigma^\nu ~=~ \Sigma^\nu e^{i k_\mu x^\mu}~~,
\eer
and attempt to obtain equations for the wave vector $k_\mu$.
If the equations have solutions which correspond to an imaginary
$k_0=\Omega$, then we conclude that our equations of motion are unstable,
since they allow for the exponential increase of a small perturbation.
After substituting the ansatz (\ref{planewave}) into (\ref{firstorder}) 
and contracting the first equation with $\Phi^A$ we obtain
\ber
\label{E1}
- k_\mu k^\mu + M_\perp^2 + 2 m_\pi^2 \Psi^0~&=&~ 
\frac{3 g }{8 \pi^2 \fpi^2} 
i k_\nu \Sigma_\mu
\eMNAB \epsilon^{ABCD} 
\Phi^A \Psi^B \partial_\alpha \Psi^C \partial_\beta \Psi^D 
\nonumber \\
\lp - k_\nu k^\nu + M^2 \rp \Sigma^\mu~&=&~
\frac{3 g }{8 \pi^2} i k_\nu \eMNAB \epsilon^{ABCD}
\Phi^A \Psi^B \partial_\alpha \Psi^C \partial_\beta \Psi^D~~.
\eer
Here we took advantage repeatedly of the perpendicularity of ${\bf \Phi}$ to 
${\bf \Psi}$. We also defined 
$M_\perp^2 = \partial_\mu \Psi^E \partial^\mu \Psi^E $ . Note that
this quantity originates in the constraint on ${\bf \Psi}$ and is not 
necessarily positive definite. 
This in itself does \em not \em imply the appearence of an instability,
since in the static case, which is free of instabilites  against small oscillations, 
$M_\perp < 0$. We should of course take the sign of $M_\perp$ into account when we
analyze the characteristic equation.

Let us solve the second equation for the polarization
vector $\Sigma^\mu$ and substitute into the first equation. 
The result can be rearranged as follows
\ber
\label{E2}
\lp - k_\mu k^\mu + M^2 \rp
\lp - k_\rho k^\rho + M_\perp^2 + 2 \mpi^2 \Psi^0 \rp~&=&~
- \lp \frac{3 g}{8 \pi^2 \fpi } \rp^2
k_\nu k_\rho ~
\eMNAB \epsilon_{\nu ~~~}^{~ \rho \sigma \tau} V_{\alpha \beta} V^{\sigma \tau}
\nonumber \\
~&=&~- \lp \frac{3 g}{8 \pi^2 \fpi } \rp^2
k_\nu k_\rho ~ S^{\mu \nu} S_{\mu ~}^{~ \rho}
\eer
where we have defined $\label{defV}
V_{\alpha \beta}~=~\epsilon^{ABCD} 
\Phi^A \Psi^B \partial_\alpha \Psi^C \partial_\beta \Psi^D$
and
$S^{\mu \nu}~=~ \eMNAB V_{\alpha\beta}$ .

\bigskip

As for any Lorentz tensor of rank 2, there are two invariants one
may construct from the components of  $S^{\mu \nu}$,
$A~=~\eMNAB S_{\mu \nu} S_{\alpha \beta}$ and
$B~=~ S^{\mu \nu}  S_{\mu \nu}$ . If $A=0$, then using the appropriate Lorentz boost,
the tensor can be either brought to a form where the 'electric' components $S^{0k}$ 
vanish or it can be brought to a form where the 'magnetic' components $S^{lm}$ vanish.
Only one of these situations is possible, depending on the sign of the other
invariant. If $B < 0$ the tensor is 'electric', and if $B > 0$ it is 'magnetic'.
Let us compute the first invariant,
\ber
\label{invA}
A~&=&~\eMNAB \epsilon_{\alpha \beta \rho \sigma} 
\epsilon_{\mu \nu \lambda \tau}
V^{\rho \sigma} V^{\lambda \tau}~=~
2 \epsilon_{\mu \nu \lambda \tau} V^{\mu \nu} V^{\lambda \tau} \nonumber \\
&=&~2 \epsilon^{ABCD} \epsilon^{EFGH} \Phi^A \Psi^B \Phi^E \Psi^F
\epsilon^{\mu \nu \alpha \beta} 
\partial_\mu \Psi^C \partial_\nu \Psi^D 
\partial_\alpha \Psi^G \partial_\beta \Psi^H~~.
\eer
 Consider one set of values for the eight isospin
indices on the right hand side. For a nonvanishing term,
the labels $\lb C, D, G, H \rb$ must be all different, otherwise we would have
symmetry in two Greek indices. For simplictiy let 
$\lb C, D, G, H \rb = \lb 0, 1, 2, 3 \rb$ . The Lorentz sum remaining to be 
performed can be rearranged,
\ber
\label{rearrange}
\epsilon_{0123} \epsilon^{\mu \nu \alpha \beta} 
\partial_\mu \Psi^0 \partial_\nu \Psi^1 \partial_\alpha \Psi^2 \partial_\beta \Psi^3 
~=~
\epsilon_{ABCD} \epsilon^{0123} 
\partial_0 \Psi^A \partial_1 \Psi^B \partial_2 \Psi^C \partial_3 \Psi^D ~=~0~~.
\eer
The term on the right hand side vanishes because it contains four isovectors
perpendicular to $\Psi$. Hence the first invariant vanishes.
The second invariant is
\ber
\label{invB}
B~&=&~S^{\mu \nu} ~S_{\mu \nu} ~=~
\epsilon^{\mu \nu \alpha \beta }
\epsilon_{\mu \nu \rho \sigma } V_{\alpha \beta} V^{\rho \sigma}~=~ 
\nonumber \\
&=&~
- 2 V_{\alpha \beta} V^{\alpha \beta} ~=~ 2 \ls V_{0k} V_{0k} - V_{lm} V_{lm} \rs~~
\nonumber \\
&=&
2 \epsilon^{ABCD} \epsilon^{EFGH} \Phi^A \Psi^B \Phi^E \Psi^F
\ls 2 \dot{\Phi}^C \partial_k \Psi^D \dot{\Phi}^G \partial_k \Psi^H -
       \partial_l \Psi^C  \partial_k \Psi^D  \partial_l \Psi^C  \partial_k \Psi^D \rs ~~.
\eer
In the last line we have lowered all Lorentz indices. 
( We remind the reader of our convention $A_\mu = (A_0,\vec{A});A^\mu = (A_0,-\vec{A})$ . )
It is clear that if the time-derivatives
are small, $B<0$ and if they are large, i.e., for fast-varying $\Psi$ fields, $B>0$. 

\bigskip

Let us consider the case with small time-derivatives. Then, $B < 0$ and the tensor $S$ may
be boosted so that $S_{lm}=0$ .
The characteristic equation is then
\ber
(M^2 - k_\mu k^\mu ) (M_\perp^2 + 2 \mpi^2 \Psi^0 - k_\nu k^\nu )~=~ - 2 {\cal C}^2 
\lb k_0 k_0 S^{l0} S^{~ 0}_{l ~} + k_l k_m S^{0l} S^{0m} \rb \nonumber \\
(M^2 + p^2 - \omega^2 ) (M_\perp^2 + 2 \mpi^2 \Psi^0 + p^2 - \omega^2 ) ~=~ 2 {\cal C}^2
\lb \omega^2 \sum\limits_l \lp S_{0l}^2 \rp - p_l p_m \lp S_{0l} S_{0m} \rp \rb~~.
\eer

Only the component of $\vec{p}$ parallel to the electric field vector $\lb E_k=S_{0k} \rb$ 
contributes to the right-hand side. Denoting that component by $p_1$, we have finally
\ber
\label{goodfinal}
(M^2 + p^2 - \omega^2 ) (M_\perp^2 + 2 \mpi^2 \Psi^0 + p^2 - \omega^2 ) ~=~ 2 {\cal C}^2  \lp  S_{0k} S_{0k} \rp
\lp \omega^2 - p_1^2 \rp~~.
\eer

When the time-derivatives dominate, we have $B > 0$ and we may choose a reference frame
where $S_{0k}=0$. The corresponding characteristic equation is
\ber
(M^2 - k_\mu k^\mu ) (M_\perp^2 + 2 \mpi^2 \Psi^0 - k_\nu k^\nu )~&=&~ - 2 {\cal C}^2 
k_l k_m S^{jl} S_{j~}^{~m} \nonumber \\
&=&~ 2 {\cal C}^2 p_l p_m S_{jl} S_{jm} ~=~ 2 {\cal C}^2  \sum\limits_j 
\lp \sum\limits_l p_l S_{jl} \rp^2 \ge 0~~.
\eer
The matrix ${\cal W}_{lm} = S_{jl} S_{jm}$ can be diagonalized 
$\tilde{\cal W}_{lm}=\delta_{lm} w_l$ and its eigenvalues 
will be real and positive as it is obvious from the preceding equation.
In that basis, the characteristic equation is
\ber
\label{badfinal}
(M^2 + p^2 - \omega^2 ) (M_\perp^2 + 2 \mpi^2 \Psi^0 + p^2 - \omega^2 ) ~=~ 2 {\cal C}^2 
\sum\limits_l p_l^2 w_l~~.
\eer

The characteristic equations in the $B<0$ and $B>0$ cases can be summarized 
as follows:
\ber
\label{goodbadshort}
(M_1^2 + p^2 - \omega^2 ) (M_2^2 + p^2 - \omega^2)~&=&~K_1^2 (\omega^2 - P_1^2 ) ~~~;~~[B<0]
\nonumber\\
(M_1^2 + p^2 - \omega^2 ) (M_2^2 + p^2 - \omega^2)~&=&~K_2^2 P_1^2  ~~~;~~[B>0]~~~.
\eer
Here, $p^2$ and $P_1^2$ are respectively the square and the square of one component of the 
arbitrary wave vector $\vec{p}=\vec{k}$ associated with the plane-wave perturbation. 
The quantities $K_1^2$ and $K_2^2$ are positive real numbers, and so is $M_1^2=M_\omega^2$.
The only exception is 
$M_2^2 = M_\perp^2+2 m_\pi^2 \Psi^0\approx M_\perp^2 = \partial_\mu \Psi^E \partial^\mu \Psi^E$.
As we have mentioned, $M_\perp^2$ is negative for a static field configuration such as a
Skyrmion at rest, which does not exhibit a proliferation of  small-wavelength perturbations. 
$M_\perp^2$  becomes positive when the time derivatives become large. 
This is the case in the center between 
the moments ($B$) and ($E$) (as defined in Sec.3), 
when most of the violent behavior takes place. It is clear from the
sign of $M_\perp^2$, as well as from the following discussion, that the characteristic equations
allow for a complex $\omega$ in a number of dynamical situations. We will focus on the regime
between the moments ($B$) and ($E$) and assume $M_\perp^2 > 0$,  and 
look for the conditions that are consistent with a negative $\omega^2$. 

In both cases ($B>0$ or $B<0$) we have a quadratic equation for $\omega^2$. 
We may solve the characteristic equation for the time constant $\omega=k_0$ using any given 
$\vec{p}$. We remind the reader the we are studying the possiblity of having a complex time
constant. Such a perturbation would grow or decrease exponentially with time. 
When the right-hand side
is small, both solutions, $\omega^2 = M_1^2, M_2^2$ are real and positive, leading to 
stable oscillations. 

In the $B<0$ case, the r.h.s. has the effect of increasing both 
the coefficient of $\omega^2$ (in absolute value) and the free term. Considering the
solutions of a quadratic equation $a X^2 - b X + c = 0$ (all coefficients are positive
here), 
$
X_{12}~=~\lp(- b \pm \sqrt{b^2 - 4 ac } \rp / 2a
$
it is clear that increasing $b$ (if both $a$ and $c$ are positive) can
make the positive real roots neither negative nor complex. Increasing $c$ actually
has the effect of bringing the roots closer together, therefore increasing the 
lower one. This may lead to trouble if the coefficient $c \ge b^2/4a $, however,
this is not possible as it is obvious if one writes the discriminant out explicitely.
We conclude that in this case, both solutions for $\omega^2$ are real and positive,
leading only to stable modes.

In the $B>0$ case there is no term containing $\omega$ on the right-hand side.
However, the free term has the opposite sign compared to the previous case. It may 
not make the discriminant negative since it decreases $c$. However, if large enough
in magnitude, this term may change the sign of $c$, thus allowing for a negative solution
for $\omega^2$ .

\begin{figure}
\centerline{
\vbox{
\hbox{
\psfig{file=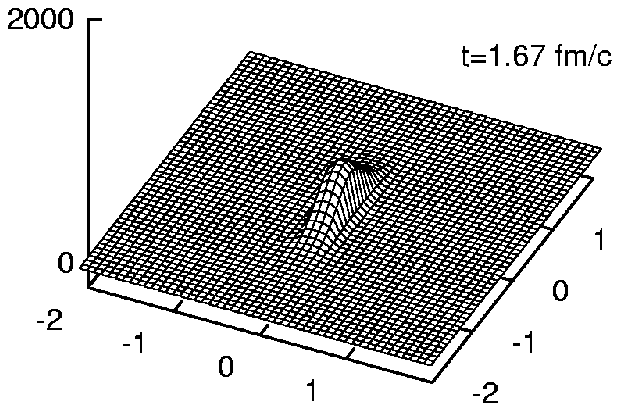,width=5.5cm,angle=0}
\psfig{file=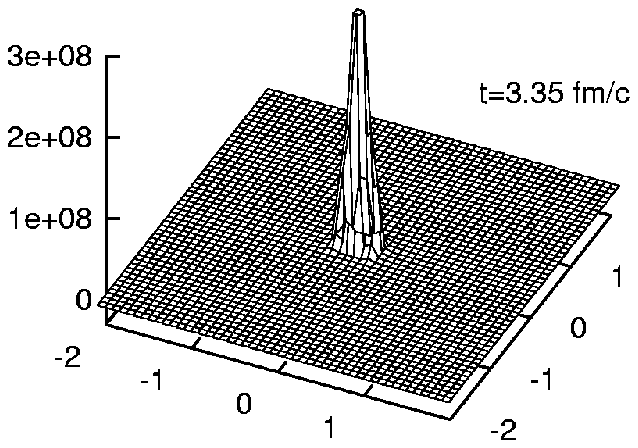,width=5.5cm,angle=0}
\psfig{file=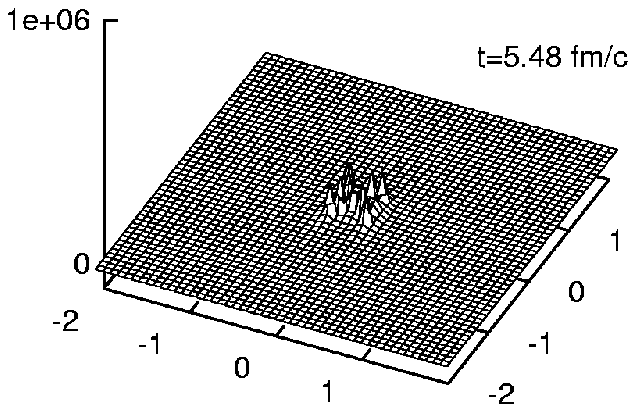,width=5.5cm,angle=0}
}
\hbox{
\psfig{file=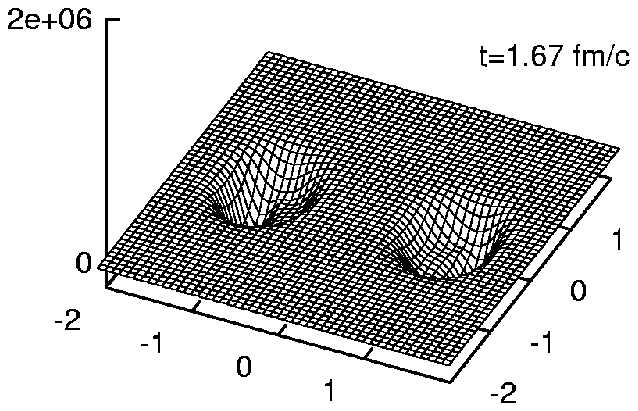,width=5.5cm,angle=0}
\psfig{file=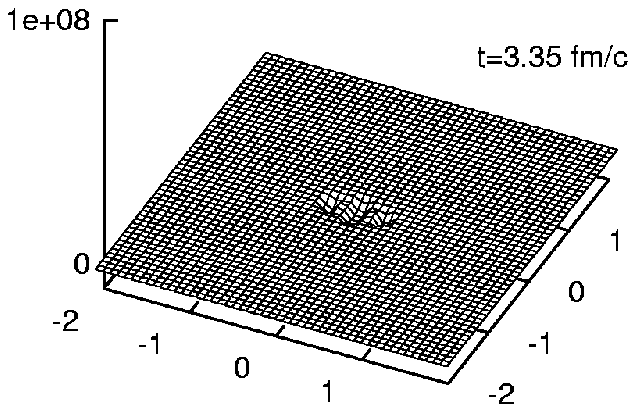,width=5.5cm,angle=0}
\psfig{file=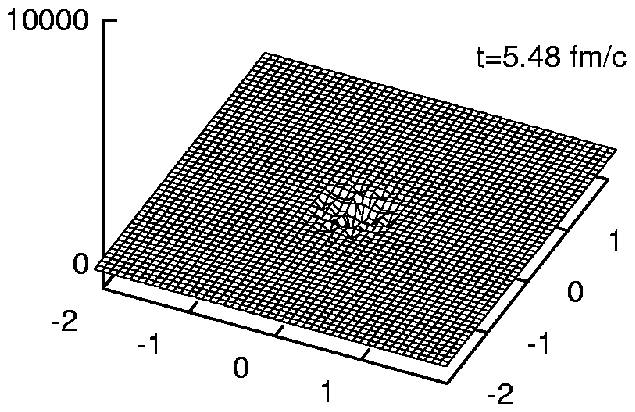,width=5.5cm,angle=0}
}
}
}
\caption{
The highest (top row) and the lowest (bottom row) eigenvalue of the tensor ${\cal T}$ at
selected moments during annihilation. 
The unit for the eigenvalues is ${\rm c^2}/ \; \fm^4$. 
Note the large differences in the vertical scales.
We plot the quantities in the $xy$ plane. The $x$ axis is the direction closer 
to horizontal. The length on the axes is measured in fermi. 
}
\label{Ev3d}
\end{figure}

\begin{figure}
\begin{center}
\psfig{file=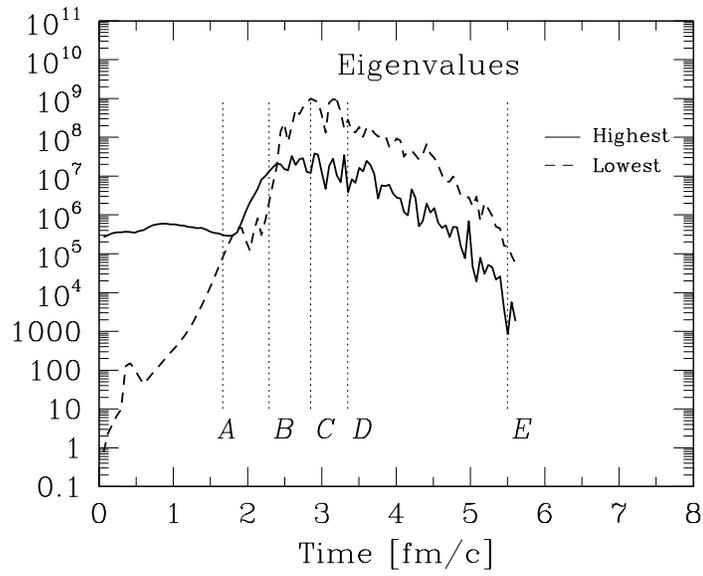,width=4.0in}
\caption{
The absolute values of the lowest (usually negative) and highest (usually positive)
eigenvalue of the tensor ${\cal T}$ overall during annihilation, in ${\rm c}^2 /\; \fm^4$. 
We associate the 
presence of large positive eigenvalues of ${\cal T}$ with instability against perturbations.
}
\label{things}
\end{center}
\end{figure}

We conclude that if $M_\perp^2 > 0$, modes with purely imaginary $\omega^2$ may occur
in the $B > 0$ case. This makes sense, because this case is associated with large time derivatives,
exactly what characterizes our violent regime. The latter is also consistent with the assumption 
$M_\perp^2 > 0$. The sign of $B$ is not immediately obvious, since it also depends on the
polarization vector of the supposed perturbation, $\Phi^A$: $B = \Phi^A \Phi^B {\cal T}^{AB}$,
where all the background-field dependent factors are contained in the tensor ${\cal T}$.
If all eigenvalues of ${\cal T}$ are negative, $B < 0$ for any polarization. If there is one
positive eigenvalue, then it is possible to have $B > 0$. If all eigenvalues are positive, 
$B > 0$ for any polarization vector. In Figure \ref{Ev3d} we plot the highest and the lowest
eigenvalues of ${\cal T}$ before, during and after the violent oscillation regime. This
illustrates that during free propagation the negative eigenvalues dominate, while during the
violent regime, the eigenvalues, and also $B$ become very large in absolute value, and the 
positive eigenvalues dominate. Towards the end of the process, the $B > 0$ regime lingers
close to the center but is not present in the outgoing radiation. 
In Figure \ref{things} we plot the absolute values of the largest and lowest eigenvalues 
of ${\cal T}$ in the whole simulation box as a function of time. The moment of half-annihilation
($B$) marks a significant increase in the magnitude of the positive eigenvalues, which dominate
in the center region through the remainder of the calculation.

In summary, the equations of motion allow in principle for the appearence of exponentially 
growing perturbations. The conditions for this are rather specific. We are able to show that
such conditions accompany the violent, fast-varying regime that follows the unwinding and 
persist until the outgoing radiation phase. However, we cannot establish a clear, causal 
connection between the $B>0$ regime and the fluctuations.

\section{Singular behavior}

The source of the persistent small oscillations is the fast-varying
behavior that follows the point of half-annihilation.
The fields at this point and shortly 
afterwards are singular, or close to that. This is the single most prominent
feature of the head-on annihilation process. The few off-center calculations
we performed show very similar features.

The head-on annihilation process has axial symmetry. 
In this case, the four 'Cartesian' components of the pion field are 
of the form
\ber
\Psi_0 (x,y,z,t) ~ = ~ f(x,\rho,t)  ~&:&~
\Psi_1 (x,y,z,t) ~ = ~ g(x,\rho,t) \nonumber \\
\Psi_2 (x,y,z,t) ~ = ~ \frac{y}{\rho} h(x,\rho,t) ~&:&~
\Psi_3 (x,y,z,t) ~ = ~ \frac{z}{\rho} h(x,\rho,t) 
\eer
where $\rho = \sqrt{y^2 + z^2}$ and the chiral constraint is $f^2 + g^2 + h^2 = 1$. 
The components of $\omega_\mu$ have a similar dependence.
The three components of $\Psi$ have additional 
symmetry constraints, namely they are all even functions of $x$. 
The radial dependence
at $\rho \rightarrow 0 $ inherits the symmetry properties of $\Psi$, $\Psi_0$ and $\Psi_1$
being even functions of $y$ and $z$ while $\Psi_2(-y) = -\Psi_2(y)$ and
 $\Psi_3(-z) = - \Psi_3 (z)$ .
Therefore $f$ and $g$ are 'even' and $h$ is and 'odd' function of $\rho$, in the sense
that $h_{|\rho=0} = 0$ and $\partial_\rho f_{|\rho = 0} =  \partial_\rho g_{|\rho = 0} = 0$
when the respective functions are continuous.

Let us consider the $\rho=0$ axis, where only $f$ and $g$ can be non-zero. 
At $x = \pm \infty$ , in free space, we have $f=1$ . As one passes through a Skyrmion,
the two components 'rotate', so that in the center of the Skyrmion we have $f=1$. 
Similarly, if we choose to move through the $\rho$ direction through the center 
of the same Skyrmion, the $g$ component would have to vanish by symmetry and the rotation
would happen in the $fh$ plane.

The moment of half-annihilation corresponds to the situation when the centers of the 
Skyrmion and the anti-Skyrmion coincide, i.e.,  $f=-1$ at $x=0$ . Now let us consider 
the radial dependence in the $x=0$ plane at this moment. At $\rho=0$ we have $f=-1$,
$g=0$, and $h=0$. At $\rho= \infty$ we have again $f=1$, $g=0$, $h=0$ . In this case 
however, the rotation happens involving mostly the $f$ and the $h$ components. At least 
close to $\rho=0$, $g$ must be much smaller than $h$, since its first $\rho$ derivative
must vanish.
\begin{figure}
\begin{center}
\psfig{file=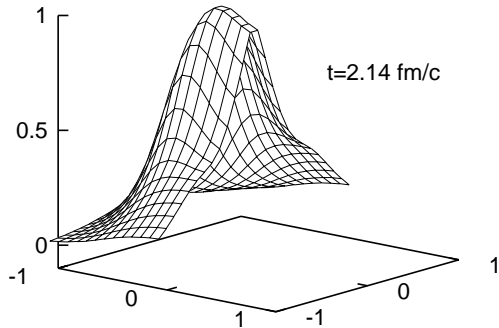,width=3.0in}
\caption{
Pion field $(1 - \Psi_0)$ close to half-annihilation. Extrapolating the field to the center 
would result in a discontinuity in the derivative.
The fields are shown in the $xy$ plane. The $x$ axis is the direction closer 
to horizontal. The length on the axes is measured in fermi. The quantity we plot is dimensionless.
}
\label{pinch}
\end{center}
\end{figure}

In Figure \ref{pinch} we plot $(1-\Psi^0(x,y))/2 = (1-f)/2$, at a moment close to half-annihilation.
The field at the center point is close to $f=-1$. As we move outwards along the $y$ (radial) 
axis, $1-f$ decreases. Notice that the variation of $f$ is is concentrated for the most part 
in a small region around $\rho=0$. 
In apparent contradiction with the requirement that $f(y)$ should be an even function,
(hence, $\partial_y f = 0$ at $y=0$), $\partial_\rho f = \partial_y f$ increases in magnitude
as $y \rightarrow 0$. The $\rho$-dependence of $f$ in Figure \ref{pinch} is well 
approximated by a $\rho^\alpha$ with $\alpha=0.2 \ldots 0.3$. 
The energy density and the baryon current are both determined by the first derivatives 
of $\Psi^A$. The effect is more dramatic in terms of these quantities. 
 \begin{figure}
\centerline{
\hbox{
\psfig{file=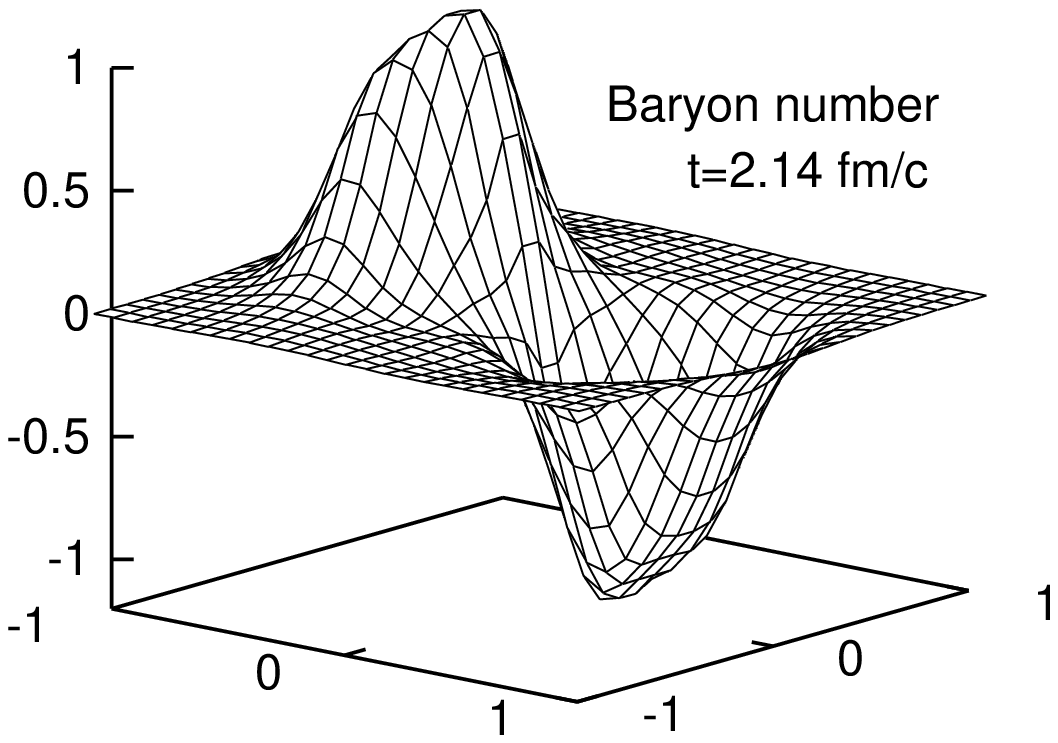,width=6cm,angle=0}
\hspace {-0.5cm}
\psfig{file=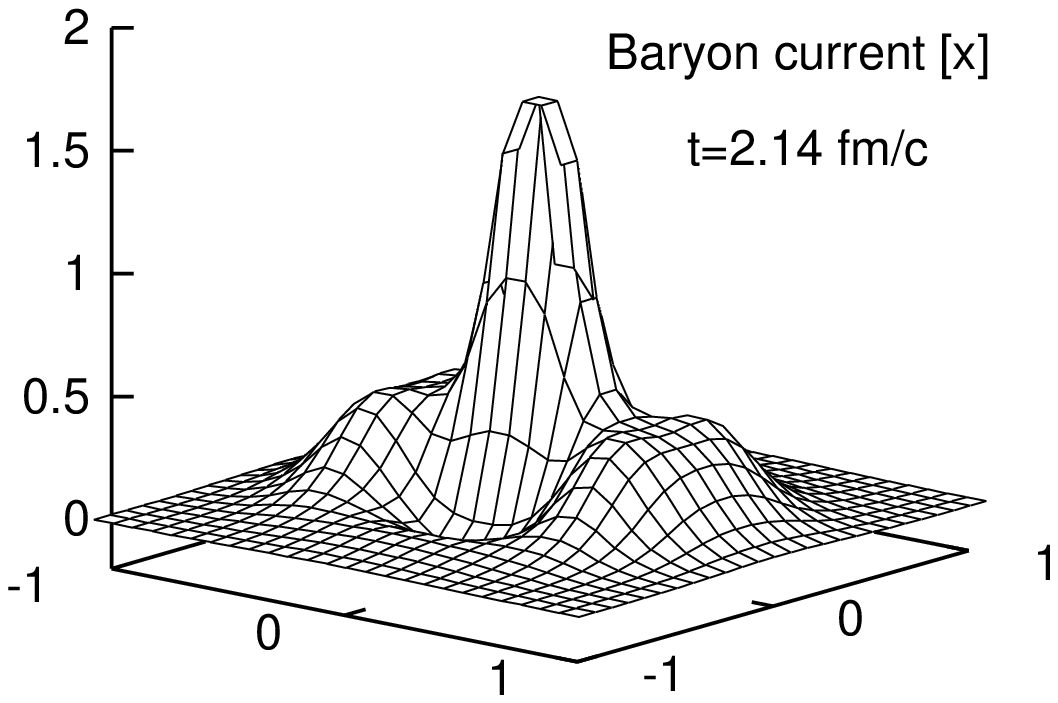,width=6cm,angle=0}
\hspace {-0.5cm}
\psfig{file=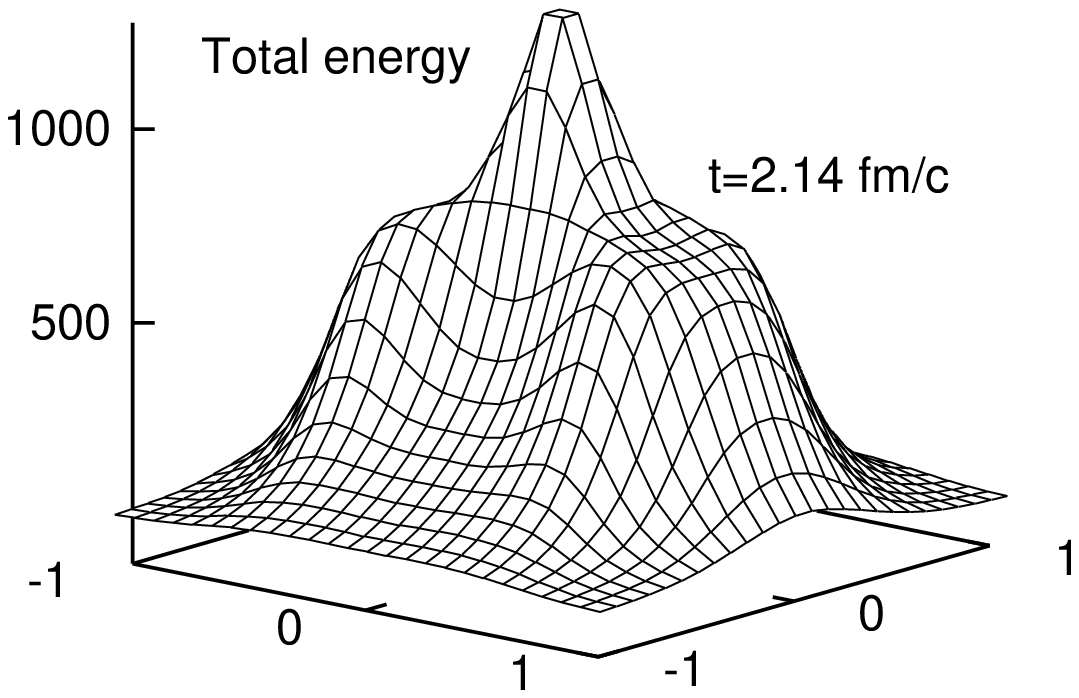,width=6cm,angle=0}
} }               
\caption{
The baryon number  density (in $\fm^{-3}$), 
the $x$ component of the baryon current density (in ${\rm c}/ \fm^{3}$),
 and the energy density ( in $\mev / \fm^3$) close to the point of half-annihilation.
The fields are shown in the $xy$ plane. The $x$ axis is the direction closer 
to horizontal. The length on the axes is measured in fermi. 
}
\label{Bpinch}
\end{figure}
In Figure \ref{Bpinch} we plot the baryon number density, the $x$ component of
the baryon current density, and the energy density in the $xy$ plane. 
The baryon current is concentrated in a very narrow region in the $x=0$ plane,
and the energy density has a steep peak that will grow dramatically during the
fast-varying regime. We wish to point out that this situation occurs at half-annihilation,
just \em before \em the messy part of the annihilation process. 
It seems very plausible that this quasi-singular configuration is the ultimate 
source of the turbulence that follows.
Because of the presence of this feature we have been forced to place our grid points so that
the symmetry center falls halfway between them in all directions, in order to avoid having
to deal with inifinite derivatives.  
We most likely miss the exact moment of the overturn  ($f=1$ at $x=\rho=0$). 
Hence we cannot tell whether the $\alpha \rightarrow 0 $ limit is achieved 
(this implies that $f(\rho)$ is a Heaviside function,
 and the baryon current is a delta function,
i.e., is all concentrated in the center).
What is clear is that as the overturn is approached the baryon current
concentrates more and more in the center, and that close to that moment, the dependence 
of the field components is consistent with a small power behavior, which in itself leads 
to infinite derivatives, hence locally infinite energy density.

A precise understanding of the singular phenomenon described above 
requires a more focused investigation. 
At this stage we have a qualitative explanation.
Recall that our $\omega$ field, apart from its mass, is an abelian
gauge field similar to the electromagnetic field.
 It couples to the respective components of
the baryon current ${\cal B}$ which, in  axial geometry, has only three independent
components, the baryon density, the $x$ component, and a radial component. 
The $\omega$ field mediated all interaction in our model.
Solitons in the pure nonlinear sigma model collapse because their energy scales as the
first power of their linear size. 
The omega field couples to the baryon (winding number) density. The fact
that this stabilizes the Skyrmions against collapse may be interpreted as a consequence of
electrostatic repulsion of the baryon charge mediated by the (omega) electric field. 

The annihilation process consists of the flow of charges of opposite signs towards the
center plane ($x=0$ in our notation), i.e., the presence of a large $B^1$ component. 
The fact that the fields vary fast in the center plane means that the baryon current is
concentrated in the center or vice versa.\footnote{For instance, if the $f$ component 
behaves like a Heaviside function,
 then the baryon current is proportional to a delta function as 
 can be seen from its expression which contains only $\rho$ and $t$ derivatives.}
 The baryon current is  very large in the center and is  pushed into a small 
cross-section in analogy with the electromagnetic 'pinch' effect enountered in plasmas.
This is a consequence of the attractive interaction between parallel electric currents.
This effect competes with the electrostatic repulsion of the charge density. 
The static charge density must vanish in the center plane, by symmetry,
(actually, all derivatives of the pion fields with respect to $x$ are zero by symmetry)
therefore here the pinch effect is strongest.
At the time of half-annihilation there is a further depletion of static charge from the 
$x=0$ region, which would explain why the pinch effect occurs first at or close half-annihilation. 
\begin{figure}
\centerline{
\hbox{
\psfig{file=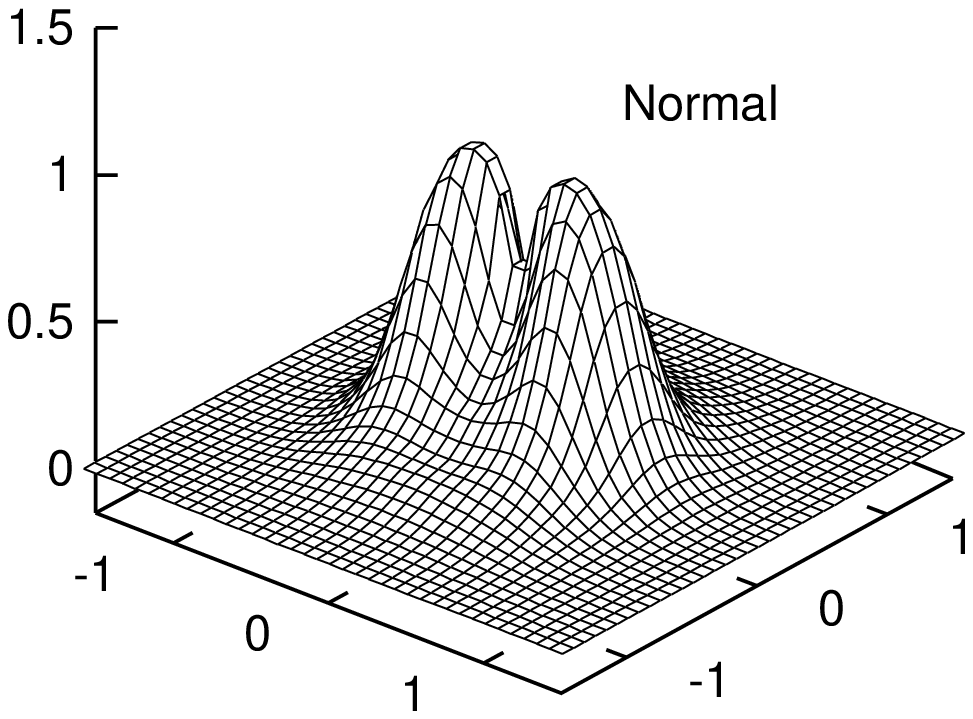,width=7.5cm,angle=0}
\hspace {0.5cm}
\psfig{file=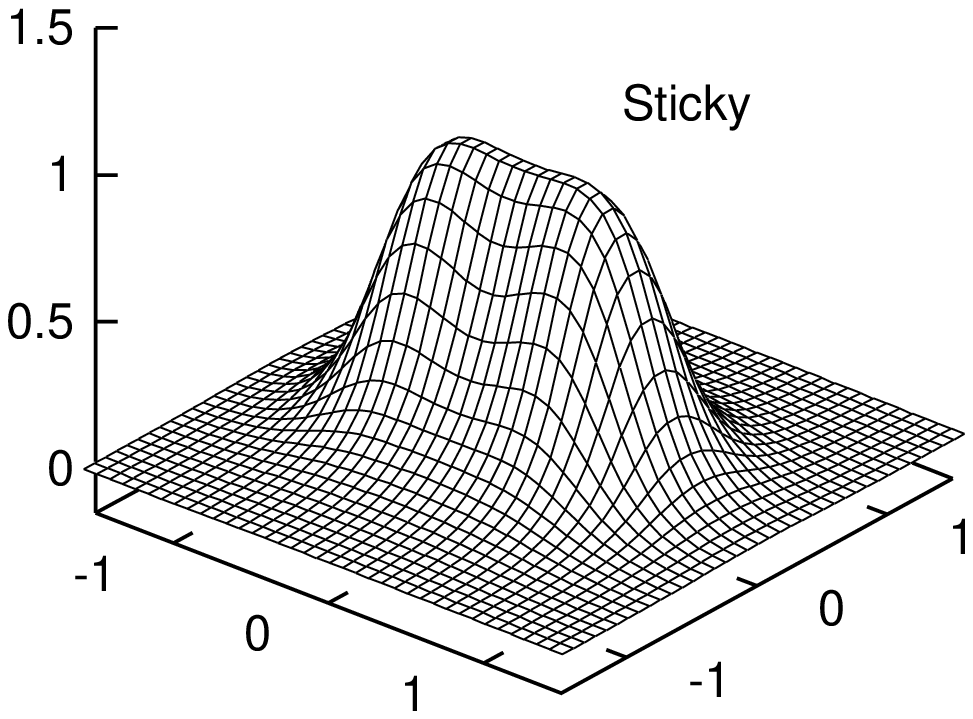,width=7.5cm,angle=0}
} }               
\caption{
One component of the pion field $(1-\Psi^0)$ close to half-annihilation, during the
true annihilation process and during a calculation where the kinetic energy is
periodically depleted by setting time derivatives to zero. The second plot shows
no pinching in the middle, indicating that the presence of a baryon current
is necessary in order to have a pinch.
The fields are shown in the $xy$ plane. The $x$ axis is the direction closer 
to horizontal. The length on the axes is measured in fermi. 
The quantity we plot is dimensionless. 
}
\label{sticky}
\end{figure}
To test whether we indeed are looking at an enectromagnetic effect, we performed a
'sticky slope' calculation, in which we periodically set the
time derivatives of the pion field, as well as the magnetic fields, to zero.
This also cancels the baryon current, but not the baryon charge, and the annihilation 
proceeds 'quasistatically'. We followed this process to the points where the baryon number
in one half space was $0.55$. In Figure \ref{sticky} we plot $1-\Psi^0$ at this moment 
in the sticky slope calculation, and, for comparison, the configuration with the same
baryon number in a calculation with the same parameters which proceeds normally. 
The pinching all but disappears in the 'quasistatic' calculation. This observation is consistent
with the contention that the baryon current is responsible for the pinching, since
in the sticky slope case there is no current, and as a result the pinching is absent as well.

While the electromagnetic effects offer a qualitative explanation, one would like to have
an approach that leads to quantitative understanding, perhaps allowing for an analytical 
description of the singular part. Furthermore, one may ask the question whether this
effect is specific to the omega-stabilized model, or is a general feature
 of dynamically stabilized solitons.

\section{Conclusion and outlook}

We have studied the classical process of annihillation of a
 Skyrmion and an anti-Skyrmion,
in a nonlinear sigma model Lagrangian which 
couples the $SU(2)$ winding number to a vector
field ($\omega$). This coupling stabilizes the Skyrmion without some of the short wave length
problems inherent in the usual fourth order Skyrme term. 
Our ultimate goal is to relate classical annihilation to
 the physical process of nucleon-antinucleon
annihillation. In this paper our goal in more modest. It is to show, for the first time,
that annihilation in the classical model can be followed numerically
 from the intial state of
separated Skyrmion and anti-Skyrmion to the final state of outgoing pion and omega 
radiation. We do encounter some violent behavior in our calculation,  but it
seems to be tamer than the fatal fluctuations previously encountered in
Skyrmion annihilation calculations \cite{Livermore,CalTech:ann}. We are able
to follow the calculations from beginning to end
 with results that are robust and independent
of the details of the turbulent behavior.

In this first attempt we only calculate Skyrmion anti-Skyrmion annihilation for 
head on collisions and only in the most attractive grooming. Our numerical code
permits other initial configurations and we plan to come back to them. We find that
annihilation happens very rapidly and is accompanied by a sharp (singular) concentration
of energy density and baryon current. This causes short wave length, noisy
 oscillations
but we are able to integrate through them. We find a burst of pion and 
omega radiation  peaked in a cone at 45 degrees with respect to the 
incident direction.  
We find turbulent behavior but the calculated outgoing meson field radiation 
carries the total incident energy to within $8 \%$. We 
show analytically that our equations 
of motion allow for a regime which is unstable with respect to the appearence of 
exponentially increasing (in time) perturbations. 
However, these conditions are met only in a spatially and temporally limited part of the 
system under study. 
This is one possible reason that  the instability does
not compromise the simulation. We find that the singular concentration of 
baryon current associated with
annihilation is analogous to the pinch effect in electromagentism. Our theory with
a vector field coupled to the current is like electromagentism, but with a mass. We
are studying ways to exploit, analytically, the nature of the singularity to
control its contribution. It would be interesting to study whether similar
singular behavior involving a peaked baryon current occurs in theories like
the standard Skyrme model, where there is no relation to electromagnetism. 

In the future we plan to study Skyrmion anti-Skyrmion collsions that are not head
on and to calculate for other groomings and 
incident energies. From this we will develop results
that can be used to extract predictions for nucleon-antinucleon physics. 
We also plan to study the singularities we encountered to see if their analytic
form can be exploited. It is sometimes the case that in the vicinity of singular behavior
one can make precise, analytic statements about solutions to problems that can
 otherwise only be addressed numerically. It will be interesting to see if that 
is the case here and to examine how general that approach is. Thus our results suggest
ample opportunities for futher work both in nucleon annihilation physics and in
mathematical physics.  

\bigskip
This work was supported in part by a grant from the United States National Science Foundation. 
The calculations have been performed mostly on the National Scalable Cluster Center
at the University of Pennsylvania. We are grateful to R. Hollebeek for his continued
support. We also thank the Eniac 2000 group for allowing us to use their resources for part of 
this work.  

We thank Randy Kamien, Pavlos Protopapas, Krishna Rajagopal, Folkert Tangerman, and 
Jac Verbaarschot for many useful discussions.

\end{document}